\documentclass[10pt, journal, twocolumn]{IEEEtran}
\usepackage[utf8]{inputenc}		
\usepackage[scr=boondoxo,frak=euler,bb=ams]{mathalfa}
\usepackage{amsmath,amsthm,amsfonts,amssymb}
\usepackage{graphicx,graphics}
\usepackage{float}
\usepackage{tikz}
\usetikzlibrary{shapes,decorations,patterns}
\usepackage{xcolor}
\newcommand{\spheading}[2][5em]{\rotatebox{90}{\parbox{#1}{\raggedright #2}}}
\usepackage[export]{adjustbox}
\usepackage[justification=centering]{caption}
\usepackage{subcaption}
\DeclareCaptionLabelSeparator{periodspace}{.\quad}
\usepackage{epstopdf}
\usepackage{enumerate,enumitem} 
\usepackage{cite}
\usepackage{suffix}
\usepackage{mathtools}
\usepackage{tabularx}
\usepackage{booktabs}
\usepackage{boldline,multirow,diagbox,hhline} 
\usepackage{authblk}
\usepackage{slashbox}
\usepackage{algorithm,algpseudocode}
\usepackage{relsize}
\usepackage{cuted}
\usepackage{titlesec}
\usepackage{chngcntr}
\theoremstyle{plain}
\newtheorem{theorem}{Theorem}[]
\newtheorem{lemma}[theorem]{Lemma} 
\newtheorem{corollary}[theorem]{Corollary}

\theoremstyle{definition} 
\newtheorem{definition}[]{Definition}
\newtheorem{remark}[]{Remark}

\newcommand{\PreserveBackslash}[1]{\let\temp=\\#1\let\\=\temp}
\newcolumntype{C}[1]{>{\PreserveBackslash\centering}p{#1}}
\newcolumntype{R}[1]{>{\PreserveBackslash\raggedleft}p{#1}} 
\newcolumntype{L}[1]{>{\PreserveBackslash\raggedright}p{#1}}
\newcolumntype{?}{!{\vrule width 1pt}}
\newcolumntype{+}{!{\vrule width 2pt}}
\DeclareFontFamily{U}{mathx}{}
\DeclareFontShape{U}{mathx}{m}{n}{<-> mathx10}{}
\DeclareSymbolFont{mathx}{U}{mathx}{m}{n}
\DeclareMathAccent{\widecheck}{0}{mathx}{"71}

\makeatletter
\newcommand\footnoteref[1]{\protected@xdef\@thefnmark{\ref{#1}}\@footnotemark}
\makeatother

\newcommand{\mbf}[1]{\boldsymbol{\mathrm{#1}}}

\newcommand{\norm}[1]{\left\lVert#1\right\rVert}
\newcommand{\abs}[1]{\left\lvert#1\right\rvert}
\newcommand{\diag}{\operatorname{diag}}

\newcommand{\summ}{\textstyle\sum\limits}

\newcommand{\psp}{\hspace{0.1em}}
\newcommand{\pspp}{\hspace{0.05em}}
\newcommand{\nsp}{\hspace{-0.1em}}
\newcommand{\nspp}{\hspace{-0.05em}}

\newcommand{\neighb}{{\rotatebox[origin=c]{90}{\footnotesize{$\rangle\nsp\langle\!$}}}_{\nsp_\mathcal{X}}\nspp}
\newcommand{\neighbb}{{\rotatebox[origin=c]{90}{\scriptsize{$\rangle\nsp\langle\!$}}}_{\nsp_\mathcal{X}}\nspp}
\newcommand{\neighbdsets}{\mathcal{D}\psp  \neighb \widecheck{\mathcal{D}}}
\newcommand{\neighbdsetss}{\mathcal{D}\psp  \neighbb \widecheck{\mathcal{D}}}
\newcommand{\supoverneighb}{\begin{array}{c} \sup\\[-1.25em] {}_{\neighbdsetss} \end{array}\!\nsp}

\newcommand{\plossrv}{\mathfrak{L}^{\mathcal{D}\nsp,\widecheck{\mathcal{D}}}_{_\mathcal{M}}}
\newcommand{\plossrvv}{\mathfrak{L}^{\widecheck{\mathcal{D}}\nsp,\mathcal{D}}_{_\mathcal{M}}}

\title{Differential Privacy with Higher Utility by Exploiting Coordinate-wise Disparity: Laplace Mechanism Can Beat Gaussian in High Dimensions}

\author{
	\IEEEauthorblockN{
		Gokularam Muthukrishnan\textsuperscript{*}, Sheetal Kalyani\textsuperscript{*}
		\thanks{\noindent\textsuperscript{*}The authors are with the Department of Electrical Engineering, Indian Institute of Technology Madras, Chennai 600036, India (e-mail: \texttt{ee17d400@smail.iitm.ac.in}; \texttt{skalyani@ee.iitm.ac.in})}}
}

\begin{document}
	
	\maketitle	
	
	\begin{abstract} 
		Conventionally, in a differentially private additive noise mechanism, independent and identically distributed (i.i.d.) noise samples are added to each coordinate of the response. In this work, we formally present the addition of noise that is independent but not identically distributed (i.n.i.d.) across the coordinates to achieve tighter privacy-accuracy trade-off by exploiting coordinate-wise disparity in privacy leakage. In particular, we study the i.n.i.d. Gaussian and Laplace mechanisms and obtain the conditions under which these mechanisms guarantee privacy. The optimal choice of parameters that ensure these conditions are derived considering (weighted) mean squared and $\ell_{\pspp p}^{\pspp p}$-errors as  measures of accuracy. Theoretical analyses and numerical simulations demonstrate that the i.n.i.d. mechanisms achieve higher utility for the given privacy requirements compared to their i.i.d. counterparts. One of the interesting observations is that the Laplace mechanism outperforms Gaussian even in high dimensions, as opposed to the popular belief, if the irregularity in coordinate-wise sensitivities is exploited. We also demonstrate how the i.n.i.d. noise can improve the performance in private (a) coordinate descent, (b) principal component analysis, and (c) deep learning with group clipping. 
	\end{abstract}
	
	\begin{IEEEkeywords}
		Differential privacy, Non-identical noise, Sensitivity profile, Gaussian mechanism, Laplace mechanism.
	\end{IEEEkeywords}
	
	\section{Introduction}
	
	\IEEEPARstart{D}{ifferential} privacy (DP) is a mathematical formulation that safeguards an individual's privacy while releasing  query responses on databases \cite{dwork2014algorithmic}. DP has become the \textit{de facto} privacy standard in machine learning applications and has been adopted in a wide range of problems such as	linear regression \cite{wang2018revisiting}, empirical risk minimization \cite{chaudhuri2011differentially}, principal component analysis \cite{dwork2014analyze}, clustering \cite{shechner2020private}, etc. Also, the US Census Bureau deployed differential privacy for the 2020 census \cite{us2021disclosure}. Differentially private  responses are necessitated to be random by definition. The parameters $\epsilon$ and $\delta\pspp$, respectively the privacy budget and privacy leakage, capture the privacy constraints. The formal definitions of DP and its mechanisms are deferred to Section \ref{sec:background}. The additive noise mechanism randomizes the result of a  query on the dataset by adding noise sampled from a known distribution and ensures privacy. When the query response is $K$-dimensional, the convention is to add independent and identically distributed (i.i.d.) noise samples to each of the coordinates; hence, the accuracy of the privatized response translates to the scale of this i.i.d. noise. Note that there is always a trade-off between privacy and utility. A stronger privacy guarantee can be achieved by adding noise of larger variance,  but this will affect the accuracy of the outcome.
	
	However, often, in a multi-dimensional query, not all the coordinates are equally vulnerable to privacy leakage.  In this article, our goal is to attain a tighter privacy-accuracy trade-off,  accounting for such irregularities. We formally present the addition of independent but non-identically distributed (i.n.i.d.) noise samples across the coordinates to ensure DP. By harnessing the underlying query-wide disparity, the i.n.i.d. noise mechanisms offer higher accuracy for the given privacy constraint than the corresponding i.i.d. mechanism.
	
	\subsection{Prior works}	
	
	Over the years, several noise distributions have been considered for differential privacy, and the privacy guarantees of such mechanisms have been documented. The Gaussian mechanism that adds i.i.d. noise from Gaussian distribution is a popular mechanism that has been studied extensively \cite{dwork2006our,dwork2014algorithmic,le2013differentially,balle2018improving,dong2022gaussian}; the Laplace mechanism is another popular mechanism that, unlike Gaussian, can ensure the stronger notion of DP with $\delta=0$. However, when employed for high dimensional queries, the i.i.d. Gaussian mechanism typically adds noise of smaller variance than the i.i.d. Laplace mechanism \cite{steinke2022composition}. This is because the $\ell_1^{}$-sensitivity, which determines the variance of Laplace noise, increases faster with the dimension than the $\ell_2^{}$-sensitivity associated with Gaussian noise.
	
	Subbotin distribution that encompasses Gaussian and Laplace as special cases has also been considered for sampling the additive noise \cite{liu2018generalized}. Recently proposed Offset Symmetric Gaussian Tails (OSGT) mechanism \cite{sadeghi2022offset} and Flipped Huber mechanism \cite{muthukrishnan2023grafting} add noise sampled from sub-Gaussian distributions and are shown to provide better accuracy than the Gaussian mechanism for the given privacy constraints. However,  obtaining the noise parameters in these mechanisms is complex, especially when the dimension of the query response is very large. Thus, Gaussian and Laplace mechanisms remain the popular choices since the scale parameter for the given privacy constraints can be determined easily. 
	
	Several noise distributions are proven to be optimal under various settings and regimes. For a single real-valued query, the optimal noise density for guaranteeing pure DP is staircase-shaped, and for a small $\epsilon$, the Laplace mechanism is optimal \cite{geng2016optimalstaircase}; for approximate DP, the truncated Laplace density \cite{geng2020tighttrunclapl} renders optimal performance in the high-privacy regime. For the one-dimensional integer-valued queries, the discrete staircase mechanism is the optimal $\epsilon$-DP mechanism \cite{geng2016optimalstaircase}; uniform and discrete Laplace distributions offer near-optimal performance under $(\epsilon,\delta)$-DP in the high-privacy regime \cite{geng2016optimaluniflap}. However, in all the aforementioned works, the emphasis was on the single-dimensional queries. In \cite{geng2015staircase}, the authors have shown that the staircase mechanism is the optimal $\epsilon$-DP mechanism that minimizes the $\ell_1^{}$-error for two-dimensional real-valued queries, where again, the Laplace noise is optimal for small $\epsilon$. To the  best of our knowledge, the optimal noise distribution in arbitrary dimensions has not been studied.
	
	Nearly all existing noise mechanisms add noise sampled from a log-concave distribution \cite{vinterbo2022differential}, as they ensure more privacy as the scale of the noise increases \cite{dong2021central}. Several works have demonstrated the central limit phenomenon in DP noise mechanisms under various conditions. In \cite{sommer2019privacy}, the authors have proven that the performance of every mechanism tends to be that of Gaussian in very high levels of composition where constituent queries are equi-sensitive. A similar result has been derived using the hypothesis testing interpretation of privacy in \cite{dong2022gaussian} for the cases where each query in the composition amounts to a small privacy leakage; in the follow-up work \cite{dong2021central}, it has been shown that the limiting behaviour is observed even for the mechanisms that add correlated noise, but with some assumptions on the homogeneity of the queries. 
	
	\subsection{Motivations}
	
	High-dimensional queries that exhibit high disparity in privacy leakage across the coordinates are very common in signal processing and machine learning applications \cite{mangold2022differentially}. The utility of the algorithms,  even the composite ones, depends on the total amount of noise added  \cite{sander2023tan}; hence, it is vital to optimize the overall amount of noise added to guarantee privacy, and leveraging the imbalance in the sensitivities to privacy leakage is an attractive option. Even if the coordinates are equally sensitive, one might perceive accuracy at certain coordinates as more important than others for the application of interest. However, such irregularities are often overlooked. Even the existing works on stronger results like optimality and central limit performance  are only for the uniform and homogeneous queries characterized by a single measure of sensitivity \cite{geng2015staircase,sommer2019privacy,dong2021central}, even when they account for noise correlation \cite{geng2015staircase,dong2021central}. It is necessary to consider the disparity intrinsic to the query in order to achieve a tighter privacy-accuracy trade-off, and failing to do so can result in higher perturbation than required	for guaranteeing privacy.
	
	A few works have accounted for such non-uniformity within the queries during noise addition. For linear queries, non-identical noise, which is also correlated, has been used to improve the accuracy \cite{li2010optimizing, nikolov2013geometry, edmonds2020power}. However, in linear queries, the underlying sensitivity structure is readily captured, which renders easy characterization of such noises. In \cite{hardt2010geometry}, the $K$-norm mechanism has been introduced in the purview of linear queries. Authors of \cite{awan2021structure} have generalized this mechanism to arbitrary queries with the introduction of sensitivity space, which captures all possible deviations observable in a query when a single user in the database is replaced. Further, they have proven that the $K$-norm mechanism characterized by the convex hull of such sensitivity space is optimal. However, characterization of the sensitivity space is not always possible; even in cases where the sensitivity space can be characterized, the construction of its convex hull and drawing noise samples are very complex for high-dimensional queries. Some of the works discuss non-identical noise addition through adaptive and uneven privacy resource allocation across the coordinates  \cite[Appx. A]{us2021disclosure}, \cite[Sec. 9]{gaboardi2016psi}, but they are based on heuristic techniques, lacking formal analysis and guarantees. In \cite{ryu2024noise}, the authors considered non-identical Laplace noise for guaranteeing per-instance DP.
	
	The choice of the noise mechanism depends on the application and its specific requirements. Gaussian and Laplace mechanisms are widely used in machine learning applications, and the tightest characterization of composition is available\footnote{Note that the definition of DP itself characterizes the composition of $\epsilon$-DP Laplace mechanism tightly, whereas the zero-concentrated DP (zCDP) offers the tightest composition result for the Gaussian mechanism \cite{bun2016concentrated,dong2022gaussian}.} to analyze multi-stage algorithms. Laplace mechanism can render strong privacy guarantees (with $\delta=0$), unlike the Gaussian mechanism. In low dimensions, Laplace outperforms Gaussian by a large margin. This can be attributed to the `sharp' centre of Laplace density, and from estimation literature, we know that the densities which are sharper tend to result in measurements that are more informative of the location parameter \cite{hogg2019introduction}; in fact, Fisher information rendered by Laplace noise is twice than that by the Gaussian noise of same variance.
	
	Asymptotic analysis suggests that  the Gaussian noise required to ensure privacy scales as  $O(\sqrt K)$ with dimension $K$, whereas the Laplace noise scales as $O(K)\pspp$, and this drives one towards the popular belief that Gaussian distribution is the ideal noise distribution for DP in high dimensions \cite{mironov2017renyi,steinke2022composition}. This asymptotic gain of Gaussian is because the Gaussian noise is calibrated with the $\ell_2^{}$-sensitivity of the query, while the scale of Laplace noise  inherently depends on  the $\ell_1^{}$-sensitivity. However, the asymptotic analyses do not present a complete picture and \textit{constants matter in differential privacy}: time and again, it has been shown that by tightening the constants, the utility can be substantially  improved  while guaranteeing the same level of privacy \cite{balle2018improving,abadi2016deep,wang2019subsampled,zhu2022optimal,asoodeh2020better}.  
	
	The above asymptotics strive to accommodate the worst-case setting where all the coordinates are equally vulnerable to privacy leakage; however, queries in real-world applications are essentially imbalanced. Thus, carefully exploiting the irregularity in the coordinate-wise sensitivities shall render the Laplace mechanism more accurate than Gaussian, alongside improving the latter's performance. Adding non-identical noise  across the coordinates is a straightforward approach for leveraging this disparity.
	
	With these as motivations, we investigate whether adding independent but non-identically distributed (i.n.i.d.) noise samples instead of i.i.d. samples provides any gain in terms of utility while guaranteeing privacy. Specifically, we introduce the i.n.i.d. variants of the Gaussian and Laplace mechanisms and provide both theoretical and empirical results to illustrate the benefits of i.n.i.d. noise addition. 
	
	\subsection{Outline of our results}
	
	\begin{enumerate}[label=(\roman*)]
		
		\item We formally present the addition of  noise that is i.n.i.d. across the coordinates of the query response so that privacy is ensured with lesser perturbation than i.i.d. noise. Introducing a new definition of sensitivity profile renders the exploitation of non-uniformity in multi-dimensional query possible. 
		
		\item In particular, we consider i.n.i.d. Gaussian and Laplace mechanisms, and the corresponding $(\epsilon,\delta)$-DP and $\epsilon$-DP guarantees are derived. The optimal choices of coordinate-wise scale parameters for these mechanisms that improve the accuracy/utility leveraging on the disparity in the coordinate-wise sensitivities  are derived. Our derivations consider the settings with (weighted) mean squared and $\ell_{\pspp p}^{\pspp p}$-errors as accuracy measures.
		
		\item Through both theoretical analyses and simulations, we show that the i.n.i.d. noise, with the proposed set of scale parameters, provides higher accuracy than i.i.d. noise. 
		
		\item Our results show that when i.n.i.d. noise is added, the Laplace mechanism can outperform Gaussian, despite ensuring the stronger $\epsilon$-DP condition, contrary to the popular belief that Gaussian noise should always be preferred in very high dimensions. For instance, when the coordinate-wise sensitivities exhibit exponential disparity, the i.n.i.d. Laplace mechanism offers lesser mean squared error (MSE) than the Gaussian for all dimensions.
		
		\item We also show that the Laplace mechanism can beat the staircase mechanism when the disparity in privacy leakage across the coordinates of a query is accounted for, even in two dimensions. 
		
		\item  We illustrate the utility of the proposed i.n.i.d. noise addition in three different applications, namely coordinate descent, principal component analysis, and deep learning with group clipping.
		
	\end{enumerate}
	
	\subsection{Basic notations} 	
	
	In this article, $\log(\cdot)$ denotes the natural logarithm. The positive part of a real number ${a}$  is denoted as $[{a }]_{\nspp +}=\max({a },0)\pspp$. $\mathbb{N}_{K}^{}$ indicates the set of first $K$ natural numbers, i.e., $\mathbb{N}_{K}^{}=\{1,\,2,\,\ldots,\,K\}$, and $\mathbb{R}_{+}^{}$ indicates the set of non-negative  real numbers, $[0,\infty)\pspp$. We use bold-face letters to denote the vectors. The operator $\norm{\cdot}_p^{}\pspp$ provides the $\ell_p^{}$-norm of a vector. The vector of all ones in $\mathbb{R}^K_{}$ is denoted as $\mbf{1}_K^{}$ and $\mbf{e}_{K,\pspp i}^{}$ is the $i$-th vector of the standard basis for $\mathbb{R}^K_{}\nspp\pspp$. We denote the Hadamard product and division by $\odot$ and $\oslash$ respectively. For the vector $\mbf{{b}}\in\mathbb{R}^K_{}\nspp\pspp$, ${b}_i^{}$ is its $i$-th entry, $\mbf{{b}}^{\circ p}_{}$ indicates $p$-th Hadamard power, $\mbf{{b}}^{\circ p}_{}=[\,{b}_1^{p}\ {b}_2^{p}\ \cdots \ {b}_K^{p}\,]^{\top}_{}$, and the diagonal matrix formed by the elements of $\mbf{{b}}$ is written as $\diag(\mbf{{b}})\pspp$. 
	
	The probability measure is denoted by $\mathbb{P}\{\cdot\}\pspp$, and $\mathbb{E}[\pspp\cdot\pspp]$ indicates the expectation operator. The probability density function (PDF) and the cumulative distribution function (CDF) of the random variable $T$ are respectively denoted as ${g}_{\pspp T}^{}(\cdot)$ and ${G}_{\pspp T}^{}(\cdot)$. The Gaussian (or normal) distribution with variance $\sigma^2$ that is centered  at ${\upsilon}$ is denoted by $\mathcal{N}({\upsilon},\sigma^2)\pspp$, and $\mathcal{L}({\upsilon},\beta)$ denotes Laplace (or bilateral exponential) distribution with mean ${\upsilon}$ and scale parameter $\beta\nspp\pspp$.  Let ${Q}(\cdot)$ denote the complementary CDF (or survival function) of the standard Gaussian distribution $\mathcal{N}(0,1)$. The notation $\Gamma(\cdot)$ is used for the gamma function,  $\Gamma({s })= \int_{0}^{{\infty}}\! {u }^{\pspp{s }-1}_{} \, e^{-{u }}_{} \, \mathrm{d}{u }\pspp$. Other notations are introduced alongside the relevant definitions in the sequel.
	
	\subsection{Organization of the paper}
	
	The rest of this article is organized as follows. Relevant background information and definitions are presented in Section \ref{sec:background}. In Section \ref{sec:inid_mech}, the scale parameters of the i.n.i.d. Gaussian and Laplace mechanisms have been derived. In Section \ref{sec:analysis}, it is theoretically shown that the i.n.i.d. mechanisms improve the accuracy of the imbalanced queries for the same level of privacy, and Section \ref{sec:empirical} numerically validates these analytical results. In Section \ref{sec:appn}, we demonstrate the efficacy of the i.n.i.d. mechanisms in three real-world applications. We discuss the strengths and limitations of the proposed mechanism in the same section. The concluding remarks are provided in Section \ref{sec:conc}. 
	
	\section{Background and i.i.d. noise mechanisms}\label{sec:background}
	
	We now provide a few definitions from differential privacy literature and introduce some additional notations that we will use in this article. In particular, we put forth a new definition of sensitivity profile, which is crucial to this work.
	
	Let $\mathcal{X}$ be the space of datasets; any dataset  $\mathcal{D} \in \mathcal{X}$ is a collection of data records from $N$ individuals. If a pair of datasets differ by only a single data record, we call them neighbouring (or adjacent) datasets; when $\mathcal{D}$ and $\widecheck{\mathcal{D}}$ are neighbouring datasets in $\mathcal{X}\pspp$, we write $\neighbdsets$. The query function ${f}:\mathcal{X}\to\mathcal{Y}$ acts on dataset $\mathcal{D}$ and outputs the query result ${f}(\mathcal{D})\in\mathcal{Y}\pspp$. The aim of DP is to conceal the presence of any individual in $\mathcal{D}$ from the query response ${f}(\mathcal{D})$ by essentially randomizing it; the algorithm $\mathcal{M}$ that provides randomized output to the query on a dataset	is known as the \textit{private mechanism}.
	
	\begin{definition}[\!\!\cite{dwork2014algorithmic}]\label{defn:DP}
		The randomized mechanism $\mathcal{M}:\mathcal{X}\to\mathcal{Y}$ is said to guarantee $(\epsilon,\delta)$-differential privacy \textup{(}$(\epsilon,\delta)$-DP in short\textup{)} if for every measurable set $\mathcal{E}$ in $\mathcal{Y}\pspp$ and every pair of neighbouring datasets $\neighbdsets$, 
		\begin{equation} \label{eq:DP}
			\mathbb{P}\{\mathcal{M}(\mathcal{D})  \in \mathcal{E}\} \leq e^{\epsilon}_{}\psp \mathbb{P}\{\mathcal{M}(\widecheck{\mathcal{D}})  \in \mathcal{E}\} + \delta	\psp,
		\end{equation}
		where $\epsilon\geq 0$ and $\delta\in[0,1]$ are respectively the privacy budget and privacy leakage parameters. When $\delta=0\pspp$, the mechanism is said to guarantee \textit{pure} DP or $\epsilon$-DP. 
	\end{definition}
	
	The notion of privacy loss encapsulates the variation between the mechanism's outputs on the neighbouring datasets as a  univariate random variable (RV). Let us denote the probability measures associated with $\mathcal{M}(\mathcal{D})$ and $\mathcal{M}(\widecheck{\mathcal{D}})$ as ${\varrho}$ and $\widecheck{{\varrho}}\pspp$, respectively. We assume that ${\varrho}$ is absolutely continuous with respect to $\widecheck{{\varrho}}$ (written as ${\varrho}\ll\widecheck{{\varrho}}\psp$), i.e., ${\varrho}$ assigns zero measure to any measurable set $\mathcal{E}\in\mathcal{Y}$ that is of zero measure under $\widecheck{{\varrho}}\pspp$, $\widecheck{{\varrho}}\pspp(\mathcal{E})=0 \implies {\varrho}\pspp(\mathcal{E})=0$  (see \cite{balle2020privacy} for generalization). The \textit{privacy loss function} is a function defined by the mapping  $\mbf{v} \mapsto \log \frac{\mathrm{d}{\varrho}}{\mathrm{d}\widecheck{{\varrho}}}(\mbf{v})$, where $\frac{\mathrm{d}{\varrho}}{\mathrm{d}\widecheck{{\varrho}}}$ is the Radon-Nikodym	derivative of ${\varrho}$ with respect to $\widecheck{{\varrho}}\pspp$, i.e., the likelihood ratio function. The random variable $\plossrv =\log\frac{\mathrm{d}{\varrho}}{\mathrm{d}\widecheck{{\varrho}}}(\mbf{{V}})\pspp$, where $\mbf{{V}} \sim {\varrho}\pspp$, is known as the \textit{privacy loss random variable} of the mechanism $\mathcal{M}$ on the neighbouring datasets $\neighbdsets$ \cite{bun2016concentrated}. The following expression (see \cite[Theorem 5]{balle2018improving}) is an equivalent condition for $(\epsilon,\delta)$-DP, which enables the interpretation of privacy guarantee through the extreme (tail) events of the privacy loss RVs:
	\begin{equation}\label{eq:balle_profile}
		\supoverneighb	\, \mathbb{P}\big\{\plossrv > \epsilon\big\}-e^{\epsilon}_{}\psp \mathbb{P}\big\{\plossrvv < -\epsilon\big\} \leq\delta	\psp;
	\end{equation}
	the term to the left of the inequality, as a function of $\epsilon$, has been named as \textit{privacy profile} \cite{balle2020privacy}. This characterization of DP through privacy losses makes the analysis easier whenever the privacy losses are sufficiently simple, as witnessed in the following subsection.
	
	\subsection{Additive noise mechanism}
	
	\begin{definition}[Additive noise mechanism]\label{defn:add_noise}
		Let ${f}: \mathcal{X} \rightarrow \mathbb{R}^{K}_{}$ be the {$K$-dimensional, real-valued} query function. The additive noise mechanism (noise mechanism, in short) imparts  differential privacy by perturbing the query output for the dataset $\mathcal{D}$ as $\mathcal{M}( \mathcal{D} ) = {f}(\mathcal{D}) + \mbf{t}	\pspp$, where $\mbf{t}=[{t}_1\ {t}_2\ \cdots\ {t}_{K}]^\top_{} \in \mathbb{R}^{K}_{}$ is the noise that is sampled from a known distribution with CDF ${G}_{\pspp\mbf{T}}^{}$ and PDF ${g}_{\pspp \mbf{T}}^{}\pspp$.
	\end{definition}
	
	Conventionally, ${t}_i,\ i \in \mathbb{N}_{K}^{} $ are i.i.d. noise samples drawn from some univariate distribution. 
	\\
	\subsubsection{Sensitivity:}
	
	The amount of noise added is determined by the privacy parameters $\epsilon$ and $\delta\pspp$. Along with these, the \textit{sensitivity} of the query function also impacts the `amount' of noise added to the true response.
	
	\begin{definition}[Sensitivity]\label{defn:sensitivity}
		For the {real-valued, $K$-dimensional} query function ${f}: \mathcal{X} \rightarrow \mathbb{R}^{K}_{}$, the $\ell_p$-sensitivity is defined as 
		\begin{equation}\label{eq:sensitivity}
			\Delta_p^{} = \supoverneighb \, \big\|{f}(\mathcal{D})-{f}(\widecheck{\mathcal{D}})\big\|_{p}^{} \,,\quad p\in[1,\infty] \psp.
		\end{equation}
		We simply denote the sensitivity as $\Delta$ when $p=\infty$ or $K=1\pspp$.
	\end{definition} 
	
	Thus, sensitivity indicates the maximum magnitude of change that the true response incurs when a single entry of the dataset is replaced. Using the equivalence of norms \cite{goldberg1987equivalence}, we have
	\begin{equation}\label{eq:norm_equiv}
		\Delta_{{q}}^{} \leq {K}^{{\left[\pspp1\nspp/\nspp {q}\pspp-\pspp1\nspp/\nspp {r}\pspp\right]_{\nspp +}^{}}}_{\stackrel{}{}}\! \times \nsp \Delta_{{r}}^{} \,,\quad \,\ \forall\, {q},\, {r} \in[1,\infty] \psp.
	\end{equation} 
	We can observe that $\Delta_{{q}}^{} \leq \Delta_{{r}}^{} \,\ \forall\, {q}\geq {r}\pspp$, and hence, $\Delta_{p}^{}$ is monotonic decreasing in ${p}\pspp$. 
	
	The main motive of this article is to improve the privacy-accuracy trade-off by leveraging the non-uniformity inherent to multi-dimensional queries. Hence, appropriate characterization of heterogeneity in sensitivity across the coordinates of a query is pivotal. The succeeding definition serves this purpose.
	\begin{definition}[Sensitivity profile]
		The \textit{sensitivity profile} is the vector of coordinate-wise sensitivities, $\mbf{{\lambda}}=[\,{\lambda}_1^{}\ {\lambda}_2^{}\ \cdots \ {\lambda}_K^{}\,]^{\top}_{}$, where 
		\begin{equation*}
			{\lambda}_i^{}=\supoverneighb \big|[{f}(\mathcal{D})-{f}(\widecheck{\mathcal{D}})]_i^{}\big|
		\end{equation*}
		is the sensitivity of the $i$-th coordinate. 
	\end{definition} 
	\begin{remark} \label{rem:assump}
		From the above definition, it is evident that $\Delta_{p}^{} \leq \norm{\mbf{{\lambda}}}_p^{}\pspp$, which holds tight for $p=\infty\pspp$. Often in literature, $\Delta_{p}^{}$ is calculated as $\norm{\mbf{{\lambda}}}_p^{}$ (for example, \cite{zhang2012functional,yu2014differentially}), as the tight computation of $\Delta_{p}^{}$ is usually difficult, especially for high-dimensional queries (see \cite[E.g. 2.1]{awan2021structure}). In this article, we will consider the approximation  $\Delta_{p}^{}=\norm{\mbf{{\lambda}}}_p^{}$ in a few instances for comparative studies, which holds true if there exists a vector $\mbf{{u}} \in \{-1,1\}^K_{}$ such that $\mbf{{u}} \odot \mbf{{\lambda}} \in \big\{\mbf{{d}}\in\mathbb{R}^K_{}\, \big\vert \, \mbf{{d}} = f(\mathcal{D}) - f(\widecheck{\mathcal{D}}),\, \neighbdsets\big\}\nspp\pspp$. This approximation is poor when the coordinates are tightly coupled (e.g., the function $f:\mathbb{N}_K^{}\to\{0,1\}^K_{}$ defined by $f(i)=\mbf{e}_{K,\pspp i}^{}\pspp$), but this is generally not the case in machine learning applications \cite{mangold2022differentially}.
	\end{remark}
	
	\subsubsection{Equivalent characterization of privacy loss:}
	
	To simplify the analysis of the additive noise mechanisms, we consider the equivalent privacy loss for the additive noise mechanism \cite{muthukrishnan2023grafting}. Let $\mbf{{z}}={f}(\mathcal{D})$ and $\widecheck{\mbf{{z}}}={f}(\widecheck{\mathcal{D}})$ denote the true responses to the query on the {neighbouring} datasets $\neighbdsets\pspp$ and let $\mbf{d}=\mbf{{z}}-\widecheck{\mbf{{z}}}$ be the difference between them. Also, let $\mbf{T}$ be the random vector that models the additive noise. Thus, the random vectors corresponding to the mechanism's outputs for $\mathcal{D}$ and $\widecheck{\mathcal{D}}$ are respectively $\mbf{{V}}=\mbf{{z}}+\mbf{T}$ and $\widecheck{\mbf{{V}}}=\widecheck{\mbf{{z}}}+\mbf{T}\pspp$.
	
	Hence, we have the output densities as ${g}_{\pspp\mbf{{V}}}^{}\nspp(\mbf{{v}})={g}_{\pspp\mbf{T}}^{}(\mbf{t})$ and ${g}_{\pspp\widecheck{\mbf{{V}}}}\nspp(\mbf{{v}})={{g}_{\pspp\mbf{T}}^{}(\mbf{t}+\mbf{d})}\pspp$, where $\mbf{t}=\mbf{{v}}-\mbf{{z}}\pspp$. The \textit{equivalent privacy loss function} is given by ${\zeta}_{\pspp \mbf{d}}^{}\nsp(\mbf{t})=\log\frac{{g}_{\pspp \mbf{T}}^{}(\mbf{t})}{{g}_{\pspp \mbf{T}}^{}(\mbf{t}+\mbf{d})}\pspp$, which is an RV that represents the privacy loss in terms of the noise density alone. It is evident that the random variable ${\zeta}_{\pspp \mbf{d}}^{}\nsp(\mbf{T})\pspp$, where $\mbf{T}\sim {G}_{\pspp\mbf{T}}^{}$ is probabilistically equivalent to  $\plossrv\pspp$. Therefore, we can express the necessary and sufficient condition for the additive noise mechanism to guarantee $(\epsilon,\delta)$-DP using the equivalent privacy losses as 
	\begin{equation}\label{eq:balle_K}
		\supoverneighb \mathbb{P}\{{\zeta}_{\pspp \mbf{d}}^{}\nsp(\mbf{T})\geq\epsilon\}-e^{\epsilon}_{}\psp \mathbb{P}\{{\zeta}_{\pspp -\mbf{d}}^{}\nsp_{}(\mbf{T})\leq-\epsilon\} \leq \delta \psp,	
	\end{equation}
	which resembles \eqref{eq:balle_profile}.   
	
	In the literature, several additive noise mechanisms have been proposed and analyzed; the prominent ones are Gaussian and Laplace, which are briefly reviewed in the sequel. 
	\\
	
	\subsubsection{Classical Gaussian mechanism:}
	
	The i.i.d. Gaussian noise of scale  $\sigma=O\nspp\Big(\nsp\tfrac{\Delta_{2}^{}}{\epsilon}\sqrt{\log\nsp\big(\tfrac{1}{\delta}\big)\!}\,\Big)\!\pspp\pspp$, when added to a query of $\ell_2^{}$-sensitivity $\Delta_2^{}\pspp$, guarantees $(\epsilon,\delta)$-DP \cite{dwork2006our,dwork2014algorithmic,bun2016concentrated}. The following result from \cite{balle2018improving} provides the privacy guarantees of the i.i.d. Gaussian mechanism.
	
	\begin{lemma}\label{lem:balle_gau}
		The Gaussian mechanism that adds i.i.d. noise sampled from $\mathcal{N}(0,\sigma^2_{})$ to each of the $K$ coordinates of the query response is $(\epsilon,\delta)$-differentially private if and only if
		\begin{equation*}
			{Q}\!\left(\nspp\tfrac{\sigma\epsilon}{\Delta_2^{}}\nspp-\nspp\tfrac{\Delta_2^{}}{2\sigma}\nspp\right)\! -e^{\epsilon}_{}  {Q}\!\left(\nspp\tfrac{\sigma\epsilon}{\Delta_2^{}}\nspp+\nspp\tfrac{\Delta_2^{}}{2\sigma}\nspp\right)\!	\psp \leq \delta \psp, 
		\end{equation*}
		where $\Delta_2^{}$ is the $\ell_2^{}$-sensitivity of the query.
	\end{lemma}
	The smallest $\sigma$ that satisfies the condition in the above lemma corresponds to the optimal i.i.d. Gaussian noise that results in the smallest perturbation of query output. Such a constant-tight scale parameter cannot be determined in closed form but can be obtained numerically \cite{balle2018improving}. 
	\\
	
	\subsubsection{Classical Laplace mechanism:}
	
	Because of the exponential tails of the noise distribution, the Laplace mechanism, unlike the Gaussian, can guarantee pure DP \cite{tian2018selective}. Under the condition given below (see \cite{dwork2006calibrating}), the i.i.d. Laplace mechanism guarantees $\epsilon$-DP.
	
	\begin{lemma}\label{lem:priv_iid_lap}
		The i.i.d. Laplace mechanism that adds $K$ independent noise samples from $\mathcal{L}(0,\beta)$ to each coordinate of the query response guarantees $\epsilon$-differentially private for $\epsilon\geq \tfrac{\Delta_1^{}}{\beta}\pspp$, where $\Delta_1^{}$ is the $\ell_1^{}$-sensitivity of the query.
	\end{lemma}
	
	Hence, the Laplace noise of scale $\tfrac{\Delta_1^{}}{\epsilon}$ corresponds to the minimum level of i.i.d. noise that is needed for $\epsilon$-DP. 
	
	\section{Non-identical noise for differential privacy} \label{sec:inid_mech}
	
	In this section, we propose to add i.n.i.d. noise that leverages  the disparity in ${\lambda}_i^{}\pspp$, $i \in \mathbb{N}_{K}^{}$ to improve the accuracy for the same privacy guarantees. The coordinate-wise scale parameters for Gaussian and Laplace noises are derived. 
	
	\subsection{Non-identical Gaussian noise mechanism}\label{sec:inid_gauss}
	
	Let us consider the DP mechanism that perturbs the query response with the noise vector whose coordinates are i.n.i.d. Gaussian random variables. The random vector $\mbf{T}=[\,T_1^{}\ T_2^{}\ \cdots \ T_K^{}\,]^{\top}_{}$ modelling the noise is multivariate Gaussian $\mathcal{N}\big(\mbf{0},(\diag(\mbf{\sigma}))^2_{}\big)\pspp$, and its coordinates $T_i\sim \mathcal{N}(0,\sigma_i^2)\pspp$, $i \in\mathbb{N}_{K}^{}$ are independent. Here, $\mbf{\sigma}=[\,\sigma_1^{}\ \sigma_2^{}\ \cdots \ \sigma_K^{}\,]^{\top}_{}\nspp\pspp$ denotes the vector of scale parameters. The scale parameters dictate the amount of noise in each coordinate, thereby  controlling the overall perturbation of the query. These are the free parameters that need to be calibrated based on the sensitivity profile of the query to guarantee a required level of privacy, and their choice is crucial  for achieving a tighter privacy-utility trade-off. 
	
	We formulate the utility-maximization problem over the set of scale parameters under the $(\epsilon,\delta)$-DP constraint. We solve this optimization problem to determine the appropriate scale parameters that result in the least perturbation utilizing the knowledge of sensitivity profile $\mbf{{\lambda}}\pspp$. We begin with the necessary and sufficient condition in terms of $\{\sigma_i^{}\}_{i=1}^K$ for the i.n.i.d. Gaussian mechanism to be $(\epsilon,\delta)$-DP.
	
	\begin{lemma}\label{lem:priv_gau_inid}
		The i.n.i.d. Gaussian mechanism that adds noise sampled from $\mathcal{N}(0,\sigma_i^2)$ to the $i$-th coordinate of the $K$-dimensional query response is $(\epsilon,\delta)$-differentially private if 
		\begin{equation}\label{eq:gauss_priv_cond}
			{Q}\!\left(\nsp\tfrac{\epsilon}{{\mu}}-\tfrac{{\mu}}{2}\nsp\right)\! -	e^{\epsilon}_{} {Q}\!\left(\nsp\tfrac{\epsilon}{{\mu}}+\tfrac{{\mu}}{2}\nsp\right)\! \psp \leq \delta \psp, 
		\end{equation} 
		where ${\mu}^2_{}={\sum_{i=1}^K \tfrac{{\lambda}_i^2}{\sigma_i^2}}$ and ${\lambda}_i^{}$ is the sensitivity of the $i$-th coordinate of the query.
	\end{lemma}
	
	\begin{proof}
		Please refer to Appendix \ref{appx:proof_sec_inid_mech}.
	\end{proof}
	
	We quantify the loss in utility brought forth by the noise using the \textit{mean squared error} (MSE). The MSE between perturbed and unperturbed query responses is related to the scale parameters as  $\mathbb{E}\nsp\big[\nsp\norm{\mathcal{M}(\mathcal{D})\nspp -\nspp{f}(\mathcal{D})}_2^2\nspp\big] =\mathbb{E}\nsp\big[\nspp\norm{\mbf{T}}_2^2\nspp\big] =\norm{\mbf{\sigma}}_{2}^2\pspp$ \cite{hogg2019introduction}. Thus, the objective is to minimize the MSE while ensuring privacy; we can obtain suitable scale parameters by solving the following optimization problem. 
	\begin{equation*}
		\text{(P1)}\quad \ \
		\begin{array}{r l}
			\min\limits_{\mbf{{\sigma}}\pspp \in\pspp \mathbb{R}^K_{+}\setminus\{\mbf 0\}} & 
			\norm{\mbf{{\sigma}}}_2^2
			\\ 
			\text{subject to}  & 
			{Q}\!\left(\nsp\tfrac{\epsilon}{{\mu}}-\tfrac{{\mu}}{2}\nsp\right)\! - e^{\epsilon}_{} {Q}\!\left(\nsp\tfrac{\epsilon}{{\mu}}+\tfrac{{\mu}}{2}\nsp\right)\! \leq\delta
			\\[1.5ex] & 
			{\mu}^2_{}={\summ_{i=1}^K \tfrac{{\lambda}_i^2}{\sigma_i^2}}
		\end{array}
		\!.
	\end{equation*} 
	
	However, the above problem is not convex, and any numerical procedure that searches for the optimum would be complex as there are $K$ parameters to be determined. Therefore, we propose to decouple the optimization into two problems. The first one deals exclusively with the privacy constraint. Let ${\mu}_0^{}$ be the maximum ${\mu}$ for which the privacy constraint holds, i.e., ${\mu}_0^{}$ is the solution to 
	\begin{equation*}
		\text{(P2)}\quad \ \
		\begin{array}{r l}
			\max\limits_{{\mu}>0} & {\mu}
			\\ 
			\text{subject to}  & 
			{Q}\!\left(\nsp\tfrac{\epsilon}{{\mu}}-\tfrac{{\mu}}{2}\nsp\right)\! - e^{\epsilon}_{} {Q}\!\left(\nsp\tfrac{\epsilon}{{\mu}}+\tfrac{{\mu}}{2}\nsp\right)\! \leq\delta
		\end{array}
		\!.
	\end{equation*} 
	Using Lemma \ref{lem:gau_priv_prof_monotonic} in Appendix \ref{appx:proof_sec_inid_mech}, we know that the constraint function is monotonically increasing in ${\mu}\nspp\pspp$. Therefore, ${\mu}_0^{}$ makes the bound in \eqref{eq:gauss_priv_cond} tighter, i.e., ${Q}\!\left(\nsp\tfrac{\epsilon}{{\mu}_0^{}}-\tfrac{{\mu}_0^{}}{2}\nsp\right)\! - e^{\epsilon}_{} {Q}\!\left(\nsp\tfrac{\epsilon}{{\mu}_0^{}}+\tfrac{{\mu}_0^{}}{2}\nsp\right)\! = \delta \pspp$, and  the privacy constraint is met by all ${\mu}\leq{\mu}_0^{}\pspp$.
	
	\begin{remark}
		Though the optimization problem (P2) is non-convex, it is one-dimensional, and hence, the  solution can be obtained using simple numerical root-finding techniques like Newton's method. From Lemma \ref{lem:gau_priv_prof_monotonic}, we know that the constraint function is monotonic: We can exploit this property to efficiently obtain the solution using the bisection method, which converges linearly\footnote{While Newton's method exhibits quadratic convergence, it  does so under  stringent conditions. We use the bisection method as it is more robust.} and finds ${\mu}_0^{}$ up to an arbitrary accuracy of ${\mu}_{\texttt{tol}}^{}$ in $O\big(\nspp\log\big({\mu}_{\texttt{tol}}^{-1}\big)\nspp\big)$  iterations \cite{burden2011numerical}. The algorithm, along with the details, has been provided in Appendix \ref{appx:bisection_method}.
	\end{remark}
	
	Once ${\mu}_0^{}$ is obtained, the optimal scale parameters can be obtained by solving the problem
	\begin{equation*}
		\text{(P3)}\quad \ \
		\begin{array}{r l}
			\min\limits_{\mbf{{\sigma}}\pspp \in\pspp \mathbb{R}^K_{+}\setminus\{\mbf 0\}} & 
			\norm{\mbf{{\sigma}}}_2^2
			\\ 
			\text{subject to}  & 
			{\summ\limits_{i=1}^K \tfrac{{\lambda}_i^2}{\sigma_i^2}}\leq {\mu}_0^{2}
		\end{array}
		\!.
	\end{equation*}
	Let us consider the function ${w}:\mathbb{R}^K_{+}\setminus\{\mbf 0\}\to\mathbb{R}$ defined by ${w}(\mbf{\sigma})=\sum_{i=1}^K \tfrac{{\lambda}_i^2}{\sigma_i^2}-{\mu}_0^{2} \pspp\pspp$. Since the Hessian matrix $\nabla^2_{\mbf{\sigma}}\, {w}(\mbf{\sigma})$ is positive definite\footnote{Since we have $\tfrac{\partial^2_{} }{\partial\sigma_j^{} \partial\sigma_l^{}}{{w}}(\mbf{\sigma})=\tfrac{6 {\lambda}_j^2 }{\sigma_j^4}\geq 0$ when $j=l$, else 0.}, ${w}$ is a convex function, and hence, the equivalent optimization problem (P3) is a convex program. 
	
	Thus, the optimal scale parameters of the problem (P1) can be obtained by solving the convex problem (P3), which in turn makes use of the solution ${\mu}_0^{}$ to the one-dimensional problem (P2). The following theorem provides the optimal i.n.i.d. noise power allocation. 
	\begin{theorem}\label{thm:opt_var_gau_inid}
		The optimal assignment of variances of the i.n.i.d. Gaussian noise that results in minimum MSE while ensuring $(\epsilon,\delta)$-DP is given by
		\begin{equation*} 
			{\sigma}_i^2=\frac{1}{{\mu}_0^2}\,{\lambda}_i^{}\norm{\mbf\lambda}_1^{}\nspp, \,\
			i \in \mathbb{N}_{K}^{} \nsp\psp,
		\end{equation*}
		where ${\mu}_0^{}$ satisfies ${Q}\!\left(\nsp\tfrac{\epsilon}{{\mu}_0^{}}-\tfrac{{\mu}_0^{}}{2}\nsp\right)\! - e^{\epsilon}_{} {Q}\!\left(\nsp\tfrac{\epsilon}{{\mu}_0^{}}+\tfrac{{\mu}_0^{}}{2}\nsp\right)\! = \delta \pspp$. 
	\end{theorem}
	
	\begin{proof}
		For generality, we determine the optimal $\mbf{\sigma}$ that minimizes the error metric $\mathbb{E}\nsp\big[\nsp\norm{\mathcal{M}(\mathcal{D})\nspp -\nspp{f}(\mathcal{D})}_{p }^{p }\nspp\big] =\mathbb{E}\nsp\big[\nspp\norm{\mbf{T}}_{p }^{p }\nspp\big] $. For the RV $T\sim \mathcal{N}(0,\sigma_{}^{2})\pspp$, we have $\mathbb{E}\big[\nsp\abs{T}^{p }_{}\nsp\big]=\sqrt{\frac{2^{p }_{}}{\pi}} \, \Gamma\big(\frac{{p }+1}{2}\big) \, \sigma_{}^{p }\pspp$, where ${p }\geq 1\pspp$. Therefore, $\mathbb{E}[\norm{\mbf{T}}^{p }_{p }] \propto \norm{\mbf{\sigma}}^{p }_{p }\pspp$, and hence, the generalized version of the problem (P3) can be written as
		\begin{equation*}
			\begin{array}{r l}
				\min\limits_{\mbf{{\sigma}}\pspp \in\pspp \mathbb{R}^K_{+}\setminus\{\mbf 0\}} & 
				\norm{\mbf{{\sigma}}}_{p }^{p }
				\\ 
				\text{subject to}  & 
				{\summ\limits_{i=1}^K \tfrac{{\lambda}_i^2}{\sigma_i^2}}\leq {\mu}_0^{2}
			\end{array}
			\!.
		\end{equation*}	
		Like (P3), this problem is also  convex. The objective function of this optimization problem has its lowest value at $\mbf{{\sigma}}=\mbf{0}$, where its gradient is zero. But, this point does not meet the constraint. Thus, the constraint is \textit{active} (i.e., the optimal solution is at the boundary of the constraint set) since the optimization is convex. Hence, the solution satisfies
		\begin{equation}\label{eq:boundary_Gauss}
			\summ\limits_{i=1}^K \tfrac{{\lambda}_i^2}{{\sigma}_i^2}={\mu}_0^2 \psp.
		\end{equation}
		We have that the objective
		\begin{equation}
			\begin{aligned}
				\norm{\mbf{{\sigma}}}_{p }^{p } 
				&= 
				\tfrac{1}{{\mu}_0^{{p }}}\big({\mu}_0^{{p }} \norm{\mbf{{\sigma}}}_{p }^{p }\big) \geq \tfrac{1}{{\mu}_0^{{p }}}\!\left(\summ_{i=1}^K \tfrac{{\lambda}_i^{2}}{{\sigma}_i^{2}}\right)^{\!\nsp \frac{{p }}{2}}_{} \!\left(\summ_{i=1}^K {{\sigma}_i^{p }}\right)\!
				\\&
				=\tfrac{1}{{\mu}_0^{{p }}}\!\left(\!\left(\summ_{i=1}^K \tfrac{{\lambda}_i^{2}}{{\sigma}_i^{2}}\right)^{\!\frac{{p }}{{p }+2}}_{}\pspp \!\left(\summ_{i=1}^K {{\sigma}_i^{p }}\right)^{\! \frac{{2 }}{p+2}}_{}\psp\right)^{\!\nspp 1+\frac{{p }}{2}}_{}\!
				\\&
				=\tfrac{1}{{\mu}_0^{{p }}}\!\left(\big\|{(\mbf{\lambda}\oslash\mbf{\sigma})^{\circ\psp \frac{2{p }}{{p }+2}}_{}}\big\|_{1+\frac{2}{{p }}}^{}\ \big\|{\mbf{\sigma}^{\circ\psp \frac{2{p }}{{p }+2}}_{}}\big\|_{1+\frac{{p }}{2}}^{} \right)^{\!\nspp 1+\frac{{p }}{2}}_{}\!
				\\&
				\geq \tfrac{1}{{\mu}_0^{{p }}}\!\left(\summ_{i=1}^K \tfrac{{\lambda}_i^{2{p }/({p }+2)}}{{\sigma}_i^{2{p }/({p }+2)}}\times{{\sigma}_i^{2{p }/({p }+2)}}\right)^{\!\nspp 1+\frac{{p }}{2}}_{}\!
				\\&
				=\tfrac{1}{{\mu}_0^{{p }}}\!\left(\summ_{i=1}^K{{\lambda}_i^{2{p }/({p }+2)}}\right)^{\!\nspp 1+\frac{{p }}{2}}_{}\!
				\psp,
			\end{aligned}
		\end{equation}
		where the first inequality is due to the privacy constraint in the problem, and the second inequality is the instantiation of H\"{o}lder's inequality\footnote{\label{fn:Holder}H\"{o}lder's inequality \cite{steele2004cauchy} states that for any vectors ${\mbf{a}}\pspp$, ${\mbf{b}} \in \mathbb{R}^K_{}$, and ${q}\pspp$, ${r} \in [1, \infty]$ satisfying $1/{q} + 1/{r}= 1$, $\norm{\mbf{a}\odot\mbf{b}}_1^{} \leq \norm{\mbf{a}}_{q}^{} \norm{\mbf{b}}_{r}^{}\pspp$, and the equality holds if and only if $\mbf{b}^{\circ {r}}_{} = \tau\pspp \mbf{a}^{\circ {q}}_{}$, for some constant $\tau\in\mathbb{R}\pspp$. The popular Cauchy-Schwarz inequality results when ${q}={r}=2\pspp$.}
		Note that the first inequality is tight for the optimal scale parameters due to \eqref{eq:boundary_Gauss}; the second is tight if
		\begin{equation*}
			\nsp\left({\sigma}_i^{2{p }/({p }+2)}\right)^{\!\nspp 1+\frac{{p }}{2}}_{}\nsp \propto \!\left(\tfrac{{\lambda}_i^{2{p }/({p }+2)}}{{\sigma}_i^{2{p }/({p }+2)}}\right)^{\!\nspp 1+\frac{2}{{p }}}_{}\! \implies {\sigma}_i^{{p }+2} \propto {\lambda}_i^{2} \psp,
		\end{equation*}
		and the proportionality constant can be determined using \eqref{eq:boundary_Gauss}. Hence, the optimal set of parameters is 
		\begin{equation}\label{eq:gauss_lp}
			{\sigma}_i^{2}=\tfrac{{\lambda}_i^{4/({p }+2)}}{{\mu}_0^2} \summ_{j=1}^K {\lambda}_j^{2{p }/({p }+2)} \pspp, \ \ i \in \mathbb{N}_{K}^{} \nspp\pspp.
		\end{equation}
		Substituting ${p }=2\pspp$, we get ${\sigma}_i^2=\tfrac{1}{{\mu}_0^2} \psp{\lambda}_i^{}\norm{\mbf{\lambda}}_1^{} \psp, \ \ i \in \mathbb{N}_{K}^{}$ as the optimal noise power distribution for the of i.n.i.d. Gaussian mechanism, resulting in the lowest MSE of $\tfrac{1}{{\mu}_0^2}\norm{\mbf{\lambda}}_1^2\pspp$. 
	\end{proof}
	
	Thus, the optimal noise variance for the $i$-th coordinate is proportional to the sensitivity of the same coordinate, $\sigma_i^2\propto {\lambda}_i^{}\pspp$. In Section \ref{sec:analysis}, we analyze the performance of the i.n.i.d. Gaussian mechanism under this optimal choice of scale parameters and its gains over the i.i.d. counterpart.
	
	\subsection{Non-identical Laplace noise mechanism}\label{sec:inid_lapl}
	
	We now introduce the i.n.i.d. Laplace mechanism that ensures $\epsilon$-DP with improved accuracy compared to the i.i.d. mechanism. Consider the random vector $\mbf{T}=[\,T_1^{}\ T_2^{}\ \cdots \ T_K^{}\,]^{\top}_{}$ whose coordinates are independent Laplace variables, $T_i^{}\sim \mathcal{L}(0,\beta_i^{})\pspp$, $i \in \mathbb{N}_{K}^{}$. The variance of $T_i^{}$ is $\sigma_i^{2}=2\beta_i^2\pspp$, and hence, the MSE resulting from the addition of i.n.i.d. Laplace noise to query output is $\norm{\mbf{\sigma}}^{2}_{2} = 2\norm{\mbf{\beta}}^{2}_{2}$. Similar to the Gaussian mechanism, the vector of scale parameters $\mbf{\beta}=[\,\beta_1^{}\ \beta_2^{}\ \cdots \ \beta_K^{}\,]^{\top}_{}\nspp\pspp$ has to be determined from the given $\epsilon$ and the coordinate-wise sensitivities ${\lambda}_i^{}, i\in \mathbb{N}_{K}^{}$. The following theorem provides the optimal choice of $\mbf{\beta}$ that minimizes the MSE.
	
	\begin{theorem}\label{thm:priv_lap_inid}
		The optimal choice of scale parameters of the $\epsilon$-differentially private i.n.i.d. Laplace mechanism adding noise sampled from $\mathcal{L}(0,\beta_i^{})$ to the $i$-th coordinate of the $K$-dimensional query response that results in minimum MSE is
		\begin{equation*}
			{\beta}_i^{}=\frac{1}{\epsilon}\, {\lambda}_i^{1/3}\,\big\Vert{\mbf{{\lambda}}_{}^{\circ \nspp\frac{2}{3}}}\big\Vert_1^{}, \,\ i \in \mathbb{N}_{K}^{}
			\nspp\pspp,
		\end{equation*} 
		where ${\lambda}_i^{}$ is the sensitivity of the $i$-th coordinate of the query. 
	\end{theorem}
	
	\begin{proof}
		The noise mechanism guarantees $\epsilon$-DP when \cite{dwork2014algorithmic}
		\begin{equation}\label{eq:pDP_cond}
			{\zeta}_{\pspp \mbf{d}}^{}\nsp(\mbf{t})=\sum\limits_{i=1}^K {\zeta}_{\pspp d_i^{}}^{}\nsp(t_i^{}) \leq \epsilon \ \,\ \forall\ \mbf{t}\in\mathbb{R}^K_{} \psp.
		\end{equation} 
		For the mechanism that adds i.n.i.d. Laplace noise, we have ${g}_{\pspp T_i}^{}(t_i)= \frac{1}{2\beta_i^{}}e^{-{|t_i^{}|}/{\beta_i^{}}}_{}$ and hence, ${\zeta}_{\pspp d_i^{}}^{}\nsp(t_i^{})=\log\tfrac{{g}_{\pspp T_i}^{}(t_i^{})}{{g}_{\pspp T_i}^{}(t_i^{}+d_i^{})} = \tfrac{|t_i^{}+d_i^{}|-|t_i^{}|}{\beta_i^{}}\pspp$. Therefore,
		\begin{align*}
			{\zeta}_{\pspp \mbf{d}}^{}\nsp(\mbf{t}) \psp = \psp \summ_{i=1}^K \tfrac{|t_i^{}+d_i^{}|-|t_i^{}|}{\beta_i^{}} \psp \leq \psp \summ_{i=1}^K \tfrac{|d_i^{}|}{\beta_i^{}} \psp \leq \psp \summ_{i=1}^K \tfrac{{\lambda}_i^{}}{\beta_i^{}} \psp,
		\end{align*}
		where the first inequality is the application of triangle inequality, and the second inequality follows from the definition of ${\lambda}_{i}^{}\pspp$. Hence, from \eqref{eq:pDP_cond}, the condition of $\epsilon$-DP is $\sum_{i=1}^K \tfrac{{\lambda}_i^{}}{\beta_i^{}} \leq \epsilon \pspp$. The MSE is $\norm{\mbf{\sigma}}^{2}_{2} = 2\norm{\mbf{\beta}}^{2}_{2}\pspp$, and the choice of scale parameters that minimize MSE while satisfying the $\epsilon$-DP constraint can be obtained by solving the optimization problem
		\begin{equation*}
			\text{(P4)}\quad \ \
			\min\limits_{\mbf{{\beta}}\pspp \in\pspp \mathbb{R}^K_{+}\setminus\{\mbf 0\}}\,   \norm{\mbf{{\beta}}}_2^2  \ \pspp \text{ subject to } \ \pspp \summ_{i=1}^K \tfrac{{\lambda}_i^{}}{\beta_i^{}} \leq \epsilon \psp,
		\end{equation*} 
		which  is a convex problem.
		
		Like the Gaussian case, we solve a generalized problem.  For ${p }\geq1$, we have $\mathbb{E}\big[\nsp\abs{T}^{p }_{}\nsp\big]=\Gamma({p+1}) \, \beta^{p }_{}$ when $T\sim \mathcal{L}(0,\beta)\pspp$, and hence, $\mathbb{E}[\norm{\mbf{T}}^{p }_{p }] \propto \norm{\mbf{\beta}}^{p }_{p }\pspp$. Thus, the optimal parameters for the generalized problem can be obtained by solving (P4), but with $\norm{\mbf{\beta}}^{p }_{p }$ as the objective function. Also, the privacy constraint associated with this convex problem is active, i.e.,
		\begin{equation}\label{eq:boundary_lap}
			\begin{array}{c}
				\sum\limits_{i=1}^K \tfrac{{\lambda}_i^{}}{{\beta}_i^{}}=\epsilon
			\end{array}\!\!\nsp
			\psp; 
		\end{equation}
		the objective is
		\begin{equation*}
			\begin{aligned}
				\norm{\mbf{{\beta}}}_{p }^{p } 
				&= 
				\tfrac{1}{\epsilon_{}^{p }}\big(\epsilon_{}^{p } \norm{\mbf{{\beta}}}_{p }^{p }\big) \geq \tfrac{1}{\epsilon_{}^{p }}\!\left(\summ_{i=1}^K \tfrac{{\lambda}_i^{}}{{\beta}_i^{}}\right)^{\!\nspp {p }}_{}\! \left(\summ_{i=1}^K {{\beta}_i^{p }}\right)\!
				\\&
				=\tfrac{1}{\epsilon_{}^{p }}\!\left(\!\left(\summ_{i=1}^K \tfrac{{\lambda}_i^{}}{{\beta}_i^{}}\right)^{\!\nspp \frac{{p }}{{p }+1}}_{}\pspp \!\left(\summ_{i=1}^K {{\beta}_i^{p }}\right)^{\!\nspp \frac{1}{{p }+1}}_{}\psp\right)^{\!\nspp {p }+1}_{}\!
				\\&
				=\tfrac{1}{\epsilon_{}^{p }}\!\left(\big\|{(\mbf{\lambda}\oslash\mbf{\beta})^{\circ\psp \frac{{p }}{{p }+1}}_{}}\big\|_{1+\frac{1}{{p }}}^{}\ \big\|{\mbf{\beta}^{\circ\psp \frac{{p }}{{p }+1}}_{}}\big\|_{{p }+1}^{} \right)^{\!\nspp {p }+1}_{}\!
				\\&
				\geq \tfrac{1}{\epsilon_{}^{p }}\!\left(\summ_{i=1}^K \tfrac{{\lambda}_i^{{p }/({p }+1)}}{{\beta}_i^{{p }/({p }+1)}}\times{{\beta}_i^{{p }/({p }+1)}}\right)^{\!\nspp {p }+1}_{}\!
				\\&
				=\tfrac{1}{\epsilon_{}^{p }}\!\left(\summ_{i=1}^K{{\lambda}_i^{{p }/({p }+1)}}\right)^{\!\nspp {p }+1}_{}\! \psp,
			\end{aligned}
		\end{equation*}
		where the first inequality arises from the privacy constraint in the problem, and it is tight for the optimal scale parameters. The second inequality is the application of H\"{o}lder's inequality\footnoteref{fn:Holder}, which is tight when
		\begin{equation*}
			\nsp\left({\beta}_i^{{p }/({p }+1)}\right)^{\nsp {p }+1}_{}\nsp \propto \!\left(\tfrac{{\lambda}_i^{{p }/({p }+1)}}{{\beta}_i^{{p }/({p }+1)}}\right)^{\! 1+\frac{1}{{p }}}_{}\! \implies {\beta}_i^{} \propto {\lambda}_i^{1/({p }+1)} \psp.
		\end{equation*}
		From the activeness of the privacy constraint \eqref{eq:boundary_lap}, the proportionality constant is determined, and the optimal scale parameters are obtained as
		\begin{equation}\label{eq:lapl_lp}
			{\beta}_i^{}=\tfrac{{\lambda}_i^{1/({p }+1)}}{\epsilon} \summ_{j=1}^K {\lambda}_j^{{p }/({p }+1)} \pspp, \ \ i \in \mathbb{N}_{K}^{} \nspp\pspp.
		\end{equation}
		Substituting ${p }=2$ provides the optimal set of parameters minimizing the MSE.
	\end{proof}
	
	Hence, the optimal choices of scale parameters are proportional to the cube root of the respective sensitivities, $\beta_i^{}\propto {\lambda}_i^{1/3}\pspp$, and the corresponding MSE is given by  $\norm{\mbf{\sigma}}_2^2=2\norm{\mbf{\beta}}_2^2=\tfrac{2}{\epsilon^2_{}}\!\left(\sum_{i=1}^K {\lambda}_i^{2/3}\right)^{\nsp 3}_{}\!\pspp$. We illustrate the reduction in MSE achieved by the i.n.i.d. Laplace mechanism for various cases of $\mbf{{\lambda}}$ through simulations in Section \ref{sec:empirical}.
	\begin{remark}
		The noise parameters in Theorems \ref{thm:opt_var_gau_inid} and \ref{thm:priv_lap_inid} render the necessary and sufficient conditions for their respective privacy guarantees when the coordinates of the query are decoupled (see Remark \ref{rem:assump}). This is because the privacy constraints in the problems (P3) and (P4) are active at these parameters (see \eqref{eq:boundary_Gauss} and \eqref{eq:boundary_lap}), and the decoupled assumption ensures that these constraints are necessary and sufficient in the first place. 
	\end{remark}
	The following result provides the scale parameters for i.n.i.d. Laplace mechanism sufficient to guarantee $(\epsilon,\delta)$-DP.
	\begin{corollary}\label{thm:priv_lap_inid_suff}
		The i.n.i.d. Laplace mechanism ensures $(\epsilon,\delta)$-DP for the set of scale parameters
		\begin{equation*}
			{\beta}_i^{}=\frac{{\lambda}_i^{1/3}\,\big\Vert{\mbf{{\lambda}}_{}^{\circ \nspp\frac{2}{3}}}\big\Vert_1^{}}{\epsilon-\log(1-\delta)}\, , \,\ i \in \mathbb{N}_{K}^{} \nspp\pspp.
		\end{equation*} 
	\end{corollary}
	
	\begin{proof}
		Please refer to Appendix \ref{appx:proof_sec_inid_mech}.
	\end{proof}
	Note that the reduction in noise scales achieved by letting $\delta>0$ is insignificant\footnote{As $\delta$ is much less than $\epsilon\pspp$, the bounds $\frac{1}{\epsilon-\log(1-\delta)} \leq \frac{1}{\epsilon+\delta} \leq \frac{1}{\epsilon}$ are close.}. Therefore, in the remainder of this article, we restrict our attention to $\epsilon$-DP i.n.i.d. Laplace mechanism.
	\begin{remark}
		Throughout the derivations, we have considered MSE to be the error metric. However, in specific settings, weighted MSE or other error measures may be of interest. The optimal scale parameters minimizing the weighted MSE with weights $\mbf{\alpha}=[\,\alpha_1^{}\ \alpha_2^{}\ \cdots \ \alpha_K^{}\,]^{\top}_{}$ are the same as those in Theorems \ref{thm:opt_var_gau_inid} and \ref{thm:priv_lap_inid}, but for the sensitivity profile $\mbf \alpha \odot \mbf \lambda\pspp$. Also, the scales minimizing $\ell_{\pspp p}^{\pspp p}$-error are given in \eqref{eq:gauss_lp} and \eqref{eq:lapl_lp}; we will occasionally use $\ell_1^{}$-error in subsequent sections. 
	\end{remark}
	
	\section{Analysis and utility} \label{sec:analysis}
	
	In this section, we prove that the proposed i.n.i.d. mechanisms guarantee privacy with improved accuracy, first with intuitive arguments, followed by formal results.
	
	An intuitive way to add non-identical noise with known coordinate-wise sensitivities is to add i.i.d. noise to the scaled query response ${f}(\mathcal{D}) \oslash \mbf{{\lambda}}$ (which would have the uniform sensitivity profile), followed by rescaling with $\mbf{{\lambda}}\pspp$. This corresponds to the following additive noise mechanism, which we term the \textit{Scale-Perturb-Rescale (SPR) mechanism}. 
	\begin{definition}[Scale-Perturb-Rescale (SPR) mechanisms]\label{defn:spr_mechanism}
		\begin{equation}\label{eq:spr_mechanism}
			\widetilde{\mathcal{M}}(\mathcal{D})=\mbf{{\lambda}}\odot({f}(\mathcal{D}) \oslash \mbf{{\lambda}}+\widetilde{\mbf{t}}_0^{}) ={f}(\mathcal{D}) +\mbf{{\lambda}}\odot\widetilde{\mbf{t}}_0^{} \pspp.
		\end{equation}
		Here, the coordinates of $\widetilde{\mbf{t}}_0^{} \in \mathbb{R}^{K}_{}$ are the independent noise samples drawn from an univariate distribution, calibrated to the sensitivity profile $\mbf{1}_{K}^{}$ and the given privacy parameters $(\epsilon,\delta)\nspp\pspp$, and $\widetilde{\mbf{t}} = \mbf{{\lambda}}\odot\widetilde{\mbf{t}}_0^{}$ is the effective i.n.i.d. noise that gets added to the query response.
	\end{definition}
	It can be observed that this mechanism is identical to the setting where an existing noise mechanism is applied in parallel to the coordinates of the query, treating each as a scalar query output. 
	
	Table \ref{tab:iid_spr_opt_inid} summarizes the scales of additive noise added (and corresponding mean squared errors) for the i.i.d. and SPR mechanisms,  along with the optimal i.n.i.d scales derived in Section \ref{sec:inid_mech}. From  Lemma \ref{lem:balle_gau}, the i.i.d. scale parameters  for $(\epsilon,\delta)$-DP Gaussian mechanism are given by $\mbf{\sigma}_0^{}=\tfrac{\Delta_2^{}}{{\mu}_0^{}}\mbf{1}_K^{}\nspp\pspp$, where ${\mu}_0^{}$ is the solution to (P2), and the corresponding MSE is $\norm{\mbf{\sigma}_0^{}}_2^2=\tfrac{K\Delta_2^2}{{\mu}_0^{2}}\nspp\pspp$. For the $\epsilon$-DP i.i.d. Laplace mechanism, the noise scales $\mbf{\beta}_0^{}=\tfrac{\Delta_1^{}}{\epsilon}\mbf{1}_K^{}\nspp\pspp$ offer the MSE of $2\norm{\mbf{\beta}_0^{}}_2^2=\tfrac{2K\Delta_1^2}{{\epsilon}_{}^{2}}\nspp\pspp$. We naturally expect the SPR mechanism to provide lesser MSE compared to its i.i.d. counterpart as it makes use of the coordinate-wise sensitivities. However, this is not the case, as we will see below. 
	
	For the SPR Gaussian mechanism, $\widetilde{\sigma}_i^{}=\tfrac{\sqrt{K}}{{\mu}_0^{}}{\lambda}_i^{}\pspp$, $i \in \mathbb{N}_{K}^{}$ are the scales of the effective noise $\widetilde{\mbf{t}}\pspp$. This results in the MSE of 
	\begin{equation*}
		\norm{\widetilde{\mbf{\sigma}}}_2^2=\tfrac{K\pspp \Vert\mbf{\lambda}\Vert_2^2}{{\mu}_0^{2}}  \geq \tfrac{ K \pspp \Delta_2^2}{{\mu}_0^{2}} = \norm{\mbf{\sigma}_0^{}}_2^2 \psp.
	\end{equation*}
	Thus, the SPR Gaussian mechanism results in a larger MSE than its i.i.d. counterpart, and at best, when the query's coordinates are decoupled, and $\Delta_2^{}$ is determined from $\mbf{\lambda}$ (see Remark \ref{rem:assump}), their MSEs are equal. For the $\epsilon$-DP SPR Laplace mechanism, the scales are $\widetilde{\beta}_i^{}=\frac{K}{\epsilon}{\lambda}_i^{}\pspp$, $i \in \mathbb{N}_{K}^{}\pspp$, which render the MSE of 
	\begin{equation*}
		2\Vert{\widetilde{\mbf{{\beta}}}}\Vert_2^{2}=\tfrac{2K^2_{}\Vert\mbf{\lambda}\Vert_2^2}{{\epsilon}_{}^{2}} \geq \tfrac{2K^2_{}\Delta_2^2}{{\epsilon}_{}^{2}} \geq \tfrac{2K\Delta_1^2}{{\epsilon}_{}^{2}}=2\Vert{\mbf{{\beta}}_0^{}}\Vert_2^2\psp,
	\end{equation*}
	which indicates that, even when the coordinates are decoupled, the SPR Laplace mechanism results in a strictly larger MSE than the corresponding i.i.d. mechanism unless the sensitivity profile is uniform, in which case its MSE is at best equal to that of the i.i.d. scheme.
	
	Thus, the SPR mechanism does not provide the optimal set of i.n.i.d. scale parameters for the given privacy guarantees and, hence, does not properly exploit the disparity in privacy leakage. Despite these issues, the SPR mechanism is the commonly used approach to add i.n.i.d. noise (for instance, in \cite{mangold2022differentially,he2022exploring}; see Section \ref{sec:appn}). Moreover, this approach limits the use of Laplace noise, as it performs worse than the corresponding i.i.d. mechanism despite the complete knowledge of coordinate-wise sensitivities. 
	
	For our optimal noise power allocation as in Theorem \ref{thm:opt_var_gau_inid}, the i.n.i.d. Gaussian noise offers the MSE $\norm{\mbf{\sigma}}_2^2=\tfrac{\Vert\mbf{\lambda}\Vert_1^2}{{\mu}_0^{2}}\nspp\pspp$. Hence, when the coordinates of the query are decoupled, the reduction in MSE compared to the i.i.d. case (using the norm equivalence) is $\tfrac{\|\mbf{\sigma}_0^{}\|_2^2}{\|\mbf{\sigma}\|_2^2}=\tfrac{K\Delta_2^2}{\Delta_1^2}\in[1,K]\pspp$. Thus, the i.n.i.d. Gaussian noise always provides lesser MSE compared to the i.i.d. noise under the decoupled setting: It can give up to $K$-fold improvement when $\mbf{{\lambda}}$ is one-hot, i.e., $\mbf{\lambda}=\Delta_2^{}\pspp\mbf{e}_{K,\pspp l}^{}$ for some $l\in \mathbb{N}_{K}^{}\pspp$. Also, the performance of i.n.i.d. noise is equivalent to that of i.i.d. noise when all the coordinates are equi-sensitive, i.e., $\mbf{{\lambda}} =\tfrac{\Delta_2^{}}{\sqrt{K}}\pspp\mbf{1}_K^{}\nspp\pspp$. This suggests that the MSE reduces with the increase in disparity of the coordinate-wise sensitivities $\lambda_i^{}\pspp$, $i \in \mathbb{N}_{K}^{}$. We formally prove this conception in the sequel. With similar arguments, we can observe that the i.n.i.d. Laplace mechanism, under the decoupled setting, with scale parameters in Theorem \ref{thm:priv_lap_inid}, always results in lesser MSE than the corresponding i.i.d. scheme.
	\begin{table}[h!] 
		\caption{Scale parameters and MSE for  $(\epsilon,\delta)$-DP Gaussian and $\epsilon$-DP Laplace mechanisms.}
		\label{tab:iid_spr_opt_inid}
		\begin{center}
			\begin{tabular}{? C{1em} | C{2.2em} ? C{6.7em} | C{6.7em} | C{6.7em} ?} 
				\clineB{3-5}{2}
				\multicolumn{1}{c}{}   &   \multicolumn{1}{c?}{}    
				&  
				\begin{tabular}{c}
					\\[-2em]
					\hspace{0.25em}\textbf{i.i.d.}
				\end{tabular} 
				&  
				\begin{tabular}{c}
					\\[-2em]
					\hspace{0.25em}\textbf{SPR i.n.i.d.}  
				\end{tabular} 
				&
				\begin{tabular}{c}
					\hspace{-0.65em}\textbf{Optimal i.n.i.d.}
					\\[-0.75em]
					\hspace{-0.65em}\textbf{(Ours)}
				\end{tabular}
				\\[1.5em] 
				\clineB{3-5}{2}
				\specialrule{0.8pt}{2pt}{0pt}
				
				\multirow{2}{*}{\spheading{\centering\textbf{Gaussian}$\quad\;\ \ $}} 
				& $\mbf{\sigma}$ 
				& $\tfrac{1}{{\mu}_0^{}}\pspp \Delta_2^{}\pspp \mbf{1}_K^{}$  
				& $\frac{1}{{\mu}_0^{}}\pspp \sqrt{K}\pspp \mbf{{\lambda}}$   
				& $\frac{1}{{\mu}_0^{}}\pspp \Vert\mbf\lambda\Vert_{1}^{\frac{1}{2}} \pspp  \mbf{{\lambda}}_{}^{\circ \nspp\frac{1}{2}}$	\\[1em] 
				\clineB{2-5}{1}
				
				& \textbf{MSE} 
				& $\frac{1}{{\mu}_0^{2}}\pspp K\pspp \Delta_2^2$ 
				& $\frac{1}{{\mu}_0^{2}}\pspp K\pspp \Vert\mbf{\lambda}\Vert_2^2$	
				& $\tfrac{1}{{\mu}_0^{2}}\pspp \Vert\mbf{\lambda}\Vert_1^2$	\\[1em] 
				
				\clineB{1-5}{2}
				\specialrule{0.8pt}{2pt}{0pt}
				
				\multirow{2}{*}{\spheading{\centering\textbf{Laplace }$\;\ \ $}} 
				& $\mbf{{\beta}}$
				& $\frac{1}{\epsilon}\pspp \Delta_1^{}\pspp  \mbf{1}_K^{}$  
				& $\frac{1}{\epsilon}\pspp K\pspp  \mbf{{\lambda}}$   
				& $\frac{1}{\epsilon}\pspp \big\Vert{\mbf{{\lambda}}_{}^{\circ \nspp\frac{2}{3}}}\big\Vert_1^{}\psp\mbf{{\lambda}}_{}^{\circ \nspp\frac{1}{3}}$	\\[1em]
				\clineB{2-5}{1}
				
				& \textbf{MSE} 
				& $\frac{2}{{\epsilon}_{}^{2}}\pspp K\pspp  \Delta_1^2$ 
				& $\frac{2}{{\epsilon}_{}^{2}}\pspp K^2_{}\pspp  \Vert\mbf{\lambda}\Vert_2^2$ 
				& $\frac{2}{\epsilon^2_{}}\pspp \big\Vert{\mbf{{\lambda}}_{}^{\circ \nspp\frac{2}{3}}}\big\Vert_1^{3}$	\\[1em] 
				\clineB{1-5}{2}			
			\end{tabular}
		\end{center}
	\end{table}
	
	It is not surprising that the SPR mechanisms cannot outperform the i.i.d. schemes as the per-coordinate noise scales depend solely on the respective sensitivities, and hence, they fail to capture the disparity across the query output. Meanwhile, in the i.i.d. mechanisms, all the noise scales are set equal, so they cannot capture the disparity either. In our i.n.i.d. mechanisms, the scale parameter of each coordinate depends on their corresponding sensitivities as well as the sensitivities of all other coordinates; thus, the mean squared error is reduced by capturing both local and global features in all of the scale parameters.
	
	\subsection{Guarantees on MSE reduction}\label{sec:mse_redn}
	
	We now formally prove that the optimal i.n.i.d scales of the Gaussian and Laplace mechanisms improve the utility with the disparity in coordinate-wise sensitivities. Firstly, we perform an asymptotic analysis to quantify the level of MSE reduction over i.i.d. mechanisms in terms of a simple measure of disparity.
	\\
	
	\subsubsection{Asymptotic analysis}
	
	Recall from Definition \ref{defn:sensitivity} that $\Delta$ denotes the $\ell_\infty^{}$-sensitivity of the query. Here, we treat  $\Delta$ as a constant and absorb $\Delta^2_{}$ terms. For the i.i.d. Gaussian mechanism guaranteeing $(\epsilon,\delta)$-DP,    $\sigma_0^{}=O\nspp\Big(\nsp\tfrac{\Delta_{2}^{}}{\epsilon}\sqrt{\log\nsp\big(\tfrac{1}{\delta}\big)\!}\,\Big)\nsp$ \cite{dwork2006our,dwork2014algorithmic,bun2016concentrated}, and hence,
	\begin{equation}\label{eq:aymp_iid_Gau}
		\text{MSE}_{\,\text{i.i.d.}}^{\pspp(\mathcal{N})} =K\sigma^2_{0} =O\nspp\Big(\nsp\tfrac{K\Delta_{2}^{2}}{\epsilon^2_{}}\log\nsp\big(\tfrac{1}{\delta}\big)\!\Big)\! =O\nspp\Big(\nsp\tfrac{K_{}^{2}}{\epsilon^2_{}}\log\nsp\big(\tfrac{1}{\delta}\big)\!\Big)\!	\psp;
	\end{equation}
	for the $\epsilon$-DP Laplace counterpart,
	\begin{equation}\label{eq:aymp_iid_Lapl}
		\text{MSE}_{\,\text{i.i.d.}}^{\pspp(\mathcal{L})} =2K\beta^2_{0} =\tfrac{2K\Delta_{1}^{2}}{\epsilon^2_{}} =O\nspp\Big(\nsp\tfrac{K_{}^{3}}{\epsilon^2_{}}\Big)\! \psp.
	\end{equation}
	The following result characterizes the asymptotic reduction in the MSE rendered by the proposed i.n.i.d. schemes. We capture the disparity in sensitivities through ${\nu}\nspp\pspp$, the ratio of the mean of the coordinate-wise sensitivities to their maximum, which is non-negligible in the asymptotic sense (i.e., ${\nu}$ is considerably smaller than 1).
	
	\begin{theorem}\label{thm:asymp}
		The i.n.i.d. Gaussian mechanism guaranteeing $(\epsilon,\delta)$-DP with scale parameters in Theorem \ref{thm:opt_var_gau_inid} results in an MSE reduction by the factor of $O({\nu}^2_{})$ over the corresponding i.i.d. Gaussian mechanism, where ${\nu}=\frac{\norm{\mbf{\lambda}}_1^{}}{K\Delta}\pspp$. Similarly, the $\epsilon$-DP i.n.i.d. Laplace mechanism with the scale parameters in Theorem \ref{thm:priv_lap_inid} reduces the MSE by the factor of $O({\nu}^2_{})$ over its i.i.d. counterpart.  
	\end{theorem}
	
	\begin{proof}
		For the $(\epsilon,\delta)$-DP i.n.i.d. Gaussian mechanism, we have
		\begin{equation*}
			\text{MSE}_{\,\text{i.n.i.d.}}^{\pspp(\mathcal{N})} =\tfrac{1}{{\mu}_0^2}\nspp\bigg(\!\summ_{i=1}^K\lambda_{i}^{}\nspp\bigg)^{\! 2}_{} =\tfrac{K^2_{}\Delta^2_{}}{M_0^2} {\nu}^{2}_{} = O\nspp\Big(\nsp{\nu}^2_{}\tfrac{K_{}^{2}}{\epsilon^2_{}}\log\nsp\big(\tfrac{1}{\delta}\big)\!\Big)\! \pspp,
		\end{equation*}
		and for the $\epsilon$-DP i.n.i.d. Laplace mechanism,
		\begin{equation*}
			\begin{split}
				\text{MSE}_{\,\text{i.n.i.d.}}^{\pspp(\mathcal{L})}
				& =\tfrac{2}{\epsilon^2_{}}\nspp\bigg(\!\summ_{i=1}^K {\lambda}_i^{2/3}\nspp\bigg)^{\! 3}_{} =\tfrac{2 K^3_{}\Delta^2_{}}{\epsilon^2_{}} \nspp\bigg(\!\tfrac{1}{K}\summ_{i=1}^K \!\Big(\frac{{\lambda}_i^{}}{\Delta}\Big)^{\nsp 2/3}_{}\bigg)^{\! 3}_{}
				\\&
				\leq \tfrac{2 K^3_{}\Delta^2_{}}{\epsilon^2_{}} \nspp\Bigg(\!\tfrac{1}{\Delta^{2/3}_{}} \!\bigg(\tfrac{1}{K}\summ_{i=1}^K{{\lambda}_i^{}}\bigg)^{\nsp 2/3}_{}\Bigg)^{\! 3}_{} = O\nspp\Big(\nsp{\nu}^2_{}\tfrac{K_{}^{3}}{\epsilon^2_{}}\Big)\! \pspp,
			\end{split}
		\end{equation*}
		where the inequality is due to Jensen. Comparing with \eqref{eq:aymp_iid_Gau} and \eqref{eq:aymp_iid_Lapl}, we observe the MSE reduction by the factor of $O({\nu}^2_{})$ for both the i.n.i.d. mechanisms.
	\end{proof}
	
	\begin{remark} \label{rem:asymp}
		The above result suggests that both the i.n.i.d. mechanisms would result in the same level of reduction of $O({\nu}^2_{})$ in the MSE for a given level of disparity ${\nu}\nspp\pspp$. However, the Laplace mechanism will offer  more reduction than $O({\nu}^2_{})$: The proof involves upper bounding the MSE of the Laplace mechanism using Jensen's inequality to get a simpler expression in terms of ${\nu}\nspp\pspp$, which is not the case with Gaussian.
	\end{remark}
	The asymptotic analysis also indicates that the dependency of the utility on the dimension $K$ vanishes when the disparity in sensitivities exhibits a power-law decay in the dimension. Specifically, when  ${\nu}=O(K^{-1}_{})\pspp$, the MSE of the i.n.i.d. Gaussian mechanism is free from the influence of the dimension $K\nspp\pspp$, and when ${\nu}=O(K^{-1/2}_{})\pspp$, the amount of noise on each coordinate does not scale with dimension; the respective conditions for the i.n.i.d. Laplace mechanism are ${\nu}=O(K^{-3/2}_{})$ and ${\nu}=O(K^{-1}_{})\pspp$, but these are conservative bounds as stated in Remark \ref{rem:asymp}.
	\\
	
	\subsubsection{Performance under various sensitivity profiles}
	
	We now provide non-asymptotic results to compare the utility of the i.n.i.d mechanisms on various sensitivity profiles. Before proceeding, we introduce the notion of majorization \cite{marshall2011inequalities}, which is  a \textit{quasi-order} on the vectors based on the relative `{spread}' of their entries. 
	
	\begin{definition}[Majorization]\label{defn:majorization}
		Consider the vectors $\mbf{{a}}\pspp$, $\mbf{{b}}\in\mathbb{R}^K_{}$ and let ${a}_{(i)}^{}$ denote the $i$-th largest entry of $\mbf{{a}}\pspp$. Then $\mbf{{b}}$ is said to majorize $\mbf{{a}}\pspp$, denoted as $\mbf{{b}} \succ \mbf{{a}}$ (or $\mbf{{a}}$ is majorized by $\mbf{{b}}\pspp$, $\mbf{{a}} \prec \mbf{{b}}$), if $\sum_{i=1}^{l}{b}_{(i)}^{} \geq \sum_{i=1}^{l} a_{(i)}^{} \,\ \forall\, l \in \mathbb{N}_{K}^{}$ with equality when $l = K$, i.e., $\sum_{i=1}^{K}b_i^{}=\sum_{i=1}^{K}a_i^{}$. 
	\end{definition}
	Intuitively, $\mbf{{b}}\succ\mbf{{a}}$ means that the entries of $\mbf{{b}}$ are more dispersed than those of $\mbf{{a}}\pspp$. We will utilize the following key result from \cite{hardy1929some} in our proofs.
	
	\begin{lemma}\label{lem:schur_cvx} 
		Consider the real-valued function ${h}_0^{}:{\mathcal{S}}\to\mathbb{R}\pspp$ (where ${\mathcal{S}}\subseteq\mathbb{R}$) and the function ${h}:{\mathcal{S}}^K_{}\to\mathbb{R}\pspp$, expressed as ${h}(\mbf{{b}})=\sum_{i=1}^{K} {h}_0^{}(b_i^{})$, $\mbf{{b}} \in {\mathcal{S}}^K_{}$. If ${h}_0^{}$ is a strictly convex function on ${\mathcal{S}}\pspp$, then ${h}$ is a strictly Schur-convex function on ${\mathcal{S}}^K_{}$; that is, if $\mbf{{b}} \succ \mbf{{a}}\pspp$ on ${\mathcal{S}}^K_{}$ and $\mbf{b}$ is not a permutation of  $\mbf{a}\pspp$, then ${h}(\mbf{{b}}) > {h}(\mbf{{a}})\pspp$. 
	\end{lemma}
	
	The following theorem formally states that for two sets of coordinate-wise sensitivities, the i.n.i.d. Gaussian noise  offers higher utility  for the one that is more spread out.
	
	\begin{theorem}\label{thm:improve_gau_inid}
		Let $\mbf{{\lambda}}$ and $\widetilde{\mbf{{\lambda}}}$ be two sets of coordinate-wise sensitivities that are not permutations of each other. If $\mbf{{\lambda}}^{\circ 2}_{} \succ \widetilde{\mbf{{\lambda}}}{}^{\circ 2}_{}\pspp$, then the mean squared error of the i.n.i.d. Gaussian mechanism corresponding to $\mbf{{\lambda}}$ is lesser than that corresponding to $\widetilde{\mbf{{\lambda}}}\pspp$, i.e.,  $\norm{\mbf{\sigma}}_2^2 < \norm{\widetilde{\mbf{\sigma}}}_2^2\pspp$.
	\end{theorem}
	
	\begin{proof}
		When $\mbf{{\lambda}}^{\circ 2}_{} \succ \widetilde{\mbf{{\lambda}}}{}^{\circ 2}_{}$, from Definition \ref{defn:majorization}, we have $\sum_{i=1}^{K}{\lambda}_{i}^2 = \sum_{i=1}^{K}\widetilde{\lambda}_{i}^{ 2}\pspp$, i.e., both $\mbf{{\lambda}}$ and $\widetilde{\mbf{{\lambda}}}$ correspond to the same $\ell_2^{}$-sensitivity, $\Delta_2^{}= \widetilde{\Delta}_2^{}\pspp$, when the coordinates are decoupled. We observe that the function ${h}_0^{}:\mathbb{R}_+^{}\to \mathbb{R}\pspp$, defined by ${h}_0^{}({r})=-\sqrt{{r}}$ for ${r}\in \mathbb{R}^{}_{+}\pspp$, is strictly convex on $\mathbb{R}^{}_{+}\pspp$. Thus, using Lemma \ref{lem:schur_cvx}, ${h}(\mbf{{b}})=-\sum_{i=1}^{K}\sqrt{{b}_i}$ is a strictly Schur-convex function on $\mathbb{R}^{K}_{+}$. We proceed further by taking $\mbf{{a}}=\widetilde{\mbf{{\lambda}}}{}^{\circ 2}_{}$ and $\mbf{{b}}=\mbf{{\lambda}}^{\circ 2}_{}$; when  $\mbf{{\lambda}}^{\circ 2}_{} \succ \widetilde{\mbf{{\lambda}}}{}^{\circ 2}_{}$ and $\mbf{{\lambda}}$ is not a  permutation of $\widetilde{\mbf{{\lambda}}}\pspp$, ${h}\big(\mbf{{\lambda}}^{\circ 2}_{}\big)\nspp > {h}\big(\widetilde{\mbf{{\lambda}}}{}^{\circ 2}_{}\big) \implies \sum_{i=1}^{K}{\lambda}_i^{} < \sum_{i=1}^{K}\widetilde{\lambda}_i^{} \implies \Vert\mbf{\lambda}\Vert_1^{} <  \Vert\widetilde{\mbf{\lambda}}\Vert_1^{}\pspp \pspp$. Hence, $\norm{\mbf{\sigma}}_2^{} =\tfrac{ \Vert\mbf{\lambda}\Vert_1^2}{{\mu}_0^{2}} < \tfrac{\Vert\widetilde{\mbf{\lambda}}\Vert_1^2}{{\mu}_0^{2}}= \norm{\widetilde{\mbf{\sigma}}}_2^{} \pspp$. 
	\end{proof}
	
	We know that $\widetilde{\mbf{{\lambda}}}{}^{\circ 2}_{}=\tfrac{{\kappa}_{}^2}{K}\mbf{1}_K^{}$ is majorized by all other $\mbf{{\lambda}}^{\circ 2}_{}$ such that $\mbf{1}_K^{\top}\big(\mbf{{\lambda}}^{\circ 2}_{}\big)={\kappa}_{}^2\pspp$; this is a direct consequence of the fact that $\tfrac{\mbf{1}_K^{}}{K}$ is majorized by every other vector in the probability simplex $\big\{\mbf{{b}}\in\mathbb{R}^K_{+}\,|\,\sum_{i=1}^K b_i^{}=1 \big\}$ \cite{marshall2011inequalities}. Hence, the uniform sensitivity profile, $ {\lambda}_i^{}=\tfrac{{\kappa}}{\sqrt{K}}\,\ \forall\,i \in \mathbb{N}_{K}^{}$, results in the maximum MSE among the profiles with the same $\ell_2^{}$-norm. 
	
	The following theorem presents a similar result for the i.n.i.d. Laplace  mechanism; we omit the proof as it is similar to that of the Gaussian case.
	
	\begin{theorem}\label{thm:improve_lapl_inid}
		Let $\mbf{{\lambda}}$ and $\widetilde{\mbf{{\lambda}}}$ be two sets of coordinate-wise sensitivities that are not permutations of each other, and let $\mbf{\beta}$ and $\widetilde{\mbf{\beta}}$ be the corresponding scale parameters for Laplace noise from Theorem \ref{thm:priv_lap_inid}. If $\mbf{{\lambda}} \succ \widetilde{\mbf{{\lambda}}}\pspp$, then $\Vert{\mbf{\beta}}\Vert_2^2 < \Vert{\widetilde{\mbf{\beta}}}\Vert_2^2\pspp$, and consequently, the mean squared error of the mechanism corresponding to $\mbf{{\lambda}}$ is lesser than that corresponding to $\widetilde{\mbf{{\lambda}}}\pspp$.
	\end{theorem}
	
	\subsection{Resource allocation perspective and composition}\label{sec:resource_allocation}
	
	So far, we have considered the scale parameters for non-identical additive noise that impart privacy to multi-dimensional queries, leveraging the knowledge of coordinate-wise sensitivities. A $K$-dimensional mechanism $\mathcal{M}:\mathcal{X}\to\mathbb{R}^{K}_{}$ can also be perceived as the (\textit{non-adaptive}) composition of one-dimensional mechanisms $\mathcal{M}^{(i)}_{}:\mathcal{X}\to\mathbb{R}$ with respective sensitivities ${\lambda}_i^{}\pspp$, $i \in \mathbb{N}_{K}^{}\nspp\pspp$, and typically in a composition, there is a \textit{privacy resource} that gets accumulated over the composition.
	
	For the Gaussian mechanism, we can consider\footnote{The factor of $1/2$ has been included for compliance with the definition of ${\eta}$-zCDP from \cite{bun2016concentrated}.} ${\eta}=\tfrac{{\mu}_0^2}{2}$ (where ${\mu}_0^{}$ is the solution to (P2)) as the privacy resource as it renders the tightest characterization of the composition of Gaussian mechanisms (see \cite[Corollaries 1 and 2]{dong2022gaussian} and  \cite[Theorem 5]{sommer2019privacy}). As a consequence of basic composition \cite{dwork2014algorithmic}, the privacy budget $\epsilon$ itself is a privacy resource for $\epsilon$-DP mechanisms. With this understanding, we can interpret ${\eta}_i^{}= \tfrac{{\lambda}_i^2}{2{\sigma}_i^2}$ and $\epsilon_i^{}= \tfrac{{\lambda}_i^{}}{{\beta}_i^{}}$ as the resource allocation for the $i$-th coordinate; the privacy constraints in the problems (P3) and (P4) can be perceived as the total resource constraints and that these constraints being active (\eqref{eq:boundary_Gauss} and \eqref{eq:boundary_lap}) suggests the \textit{full utilization} of the available resources.
	
	From Theorems \ref{thm:opt_var_gau_inid} and \ref{thm:priv_lap_inid}, the  optimal distribution of resources across the coordinates for Gaussian and Laplace mechanisms are respectively 
	\begin{equation*}
		{\eta}_i^{}=\tfrac{{\lambda}_i^{}}{\sum_{j=1}^K {\lambda}_j^{}}\,{\eta}
		\ \text{ and } \
		{\epsilon}_i^{}=\tfrac{{\lambda}_i^{2/3}}{\sum_{j=1}^K {\lambda}_j^{2/3}}\,{\epsilon}
		\pspp, \ i 
		\in \mathbb{N}_{K}^{}
		\psp ;	
	\end{equation*}
	thus, optimal i.n.i.d. mechanism allocates privacy resources to the coordinates depending on the sensitivities (${\eta}\propto {\lambda}_i^{}$ and  ${\epsilon}_i^{}\propto {\lambda}_i^{2/3}$)$\pspp$. It can  be observed that the SPR scale parameters also result in active privacy constraints. However, the SPR mechanism distributes the privacy resource equally across the coordinates (i.e., $\eta_i^{}={\eta}/{K}$ and $\epsilon_i^{}={\epsilon}/{K}$); this allocation does not account for the disparity in the coordinate-wise sensitivities, and hence, the SPR mechanism is sub-optimal.
	
	\begin{figure*}[h!]
		\begin{subfigure}{0.5\linewidth}
			\centering
			\includegraphics[width=0.84\linewidth]{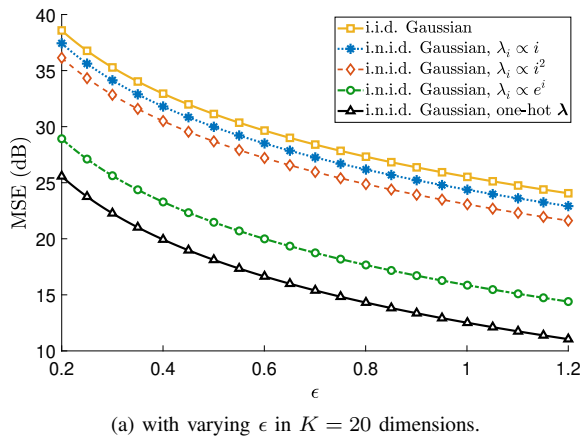}
			\caption{with varying $\epsilon$ in $K=20$ dimensions.}
			\label{fig:inid_gauss_eps_vs_mse}
		\end{subfigure}
		\begin{subfigure}{0.5\linewidth}
			\centering
			\includegraphics[width=0.84\linewidth]{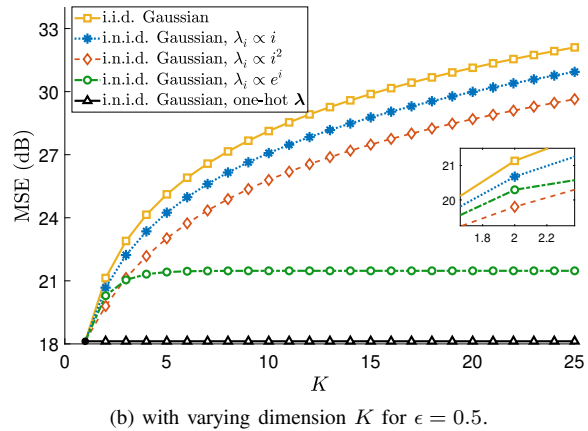}
			\caption{with varying dimension $K$ for $\epsilon=0.5\pspp$.}
			\label{fig:inid_gauss_K_vs_mse}
		\end{subfigure}
		\caption{Performance of i.i.d. and i.n.i.d. $(\epsilon,\delta)$-DP Gaussian mechanisms under various sensitivity profiles with unit $\ell_2^{}$-sensitivity for $\delta=10^{-6}_{}$.}
		\label{fig:inid_gauss}
	\end{figure*}
	
	\begin{remark}
		This resource allocation perspective is vital to broaden the scope of the non-identical noise addition to composite queries. For instance, any differentially private iterative algorithm consumes privacy resources over each iteration, and with this perspective, one can add noise of different scales across the iterations. We will witness such use cases in the applications presented in Section \ref{sec:appn}. Since our i.n.i.d. noise schemes guarantee privacy with a lesser total amount of noise, they can improve the utility of composite algorithms \cite{sander2023tan} as long as one can come up with some meaningful $\mbf{\lambda}\pspp$.
	\end{remark}
	
	\section{Empirical validation}\label{sec:empirical}
	In this section, through numerical simulations, we demonstrate the performance gains of the proposed i.n.i.d. Gaussian and Laplace noise over their i.i.d. counterparts with the increase in the level of disparity of coordinate-wise sensitivities. The \textit{theoretical} mean squared error achieved by the optimal scale parameters obtained in Theorems \ref{thm:opt_var_gau_inid} and \ref{thm:priv_lap_inid} is used as the benchmark utility metric for comparison.
	
	As illustrative examples, we consider a few simple sensitivity profiles with varying levels of disparity; we study the  performance with three different cases of coordinate-wise sensitivities, ${\lambda}_i^{}\propto i\pspp$, ${\lambda}_i^{}\propto i^2_{}\pspp$, and ${\lambda}_i^{}\propto e^i_{}\nspp\pspp$ (we call these respectively linear, quadratic, and exponential profiles); in realistic settings, we will encounter more unstructured, yet considerably disparate, sensitivity profiles as in the applications shown in Section \ref{sec:appn}. Along with these three profiles, we include the results corresponding to the edge cases of uniform and one-hot $\mbf{{\lambda}}$ to gauge the highest and lowest MSE achievable by the proposed noise parameters, disregarding their practical irrelevance. For all the cases, $\mbf{{\lambda}}$ is normalized so that $ \Vert\mbf{\lambda}\Vert_2^{}=1$ for the Gaussian mechanism, and for Laplace, $\mbf{{\lambda}}$ is scaled such that $ \Vert\mbf{\lambda}\Vert_1^{}=1\pspp$. Throughout the section, we assume that  the global sensitivities are determined from $\mbf{\lambda}$ as $\Delta_p^{}=\norm{\mbf{\lambda}}_p^{}$ (see Remark \ref{rem:assump}). We quantify the level of dispersion in $\mbf{{\lambda}}\pspp$ using the Gini coefficient \cite{marshall2011inequalities}, computed as $\frac{1}{2K\norm{\mbf{\lambda}}_1^{}}\sum_{i=1}^{K}\sum_{j=1}^{K} | {\lambda}_{i}^{}-{\lambda}_{j}^{}|\pspp$.
	
	\begin{figure}[ht]
		\centering
		\includegraphics[width=0.84\linewidth]{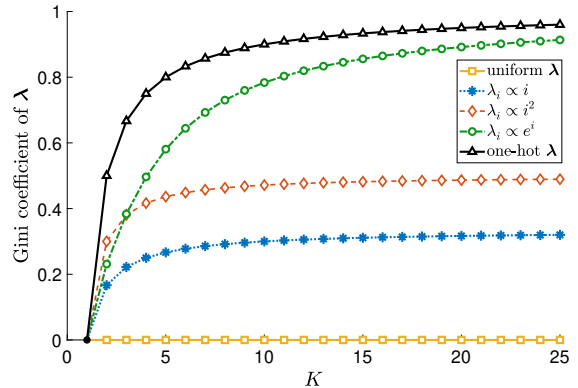}
		\caption{Gini coefficients of various sensitivity profiles having unit $\ell_2^{}$-sensitivity with varying dimension $K$.}
		\label{fig:inid_gauss_gini}
	\end{figure}
	
	\subsection{Gaussian mechanism}
	First, we analyze the MSE corresponding to i.n.i.d. and i.i.d. Gaussian mechanisms with varying privacy budget $\epsilon$ in 20 dimensions when $\delta=10^{-6}_{}\nspp\pspp$. The corresponding results are provided in Figure \ref{fig:inid_gauss_eps_vs_mse}. As the i.i.d. mechanism does not account for individual sensitivities ${\lambda}_i^{}\pspp$, the MSE remains the same irrespective of how the elements of $\mbf{{\lambda}}$ are spread. However, the i.n.i.d. noise always results in lesser MSE than the i.i.d. case. In particular, the reduction in MSE over the i.i.d. mechanism is $1.145\psp\mathrm{dB}\pspp$, $2.442\psp\mathrm{dB}\pspp$, and $9.658\psp\mathrm{dB}$ (i.e., by a factor of $1.3016\pspp$, $1.7547$ and $9.2423$), respectively, for the cases of linear, quadratic and exponential profiles and the maximum possible reduction, achievable when $\mbf{{\lambda}}$ is one-hot, is $10\log_{10}^{}(K)=13.01\psp\mathrm{dB}\pspp$.
	
	\begin{figure*}[h!]
		\begin{subfigure}{0.5\linewidth}
			\centering
			\includegraphics[width=0.84\linewidth]{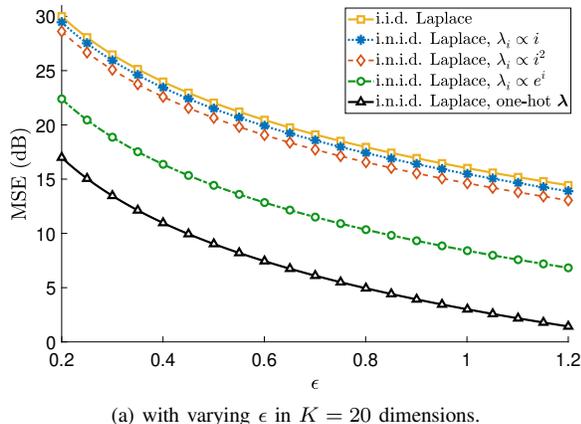}
			\caption{with varying $\epsilon$ in $K=20$ dimensions.}
			\label{fig:inid_lapl_eps_vs_mse}
		\end{subfigure}
		\begin{subfigure}{0.5\linewidth}
			\centering
			\includegraphics[width=0.84\linewidth]{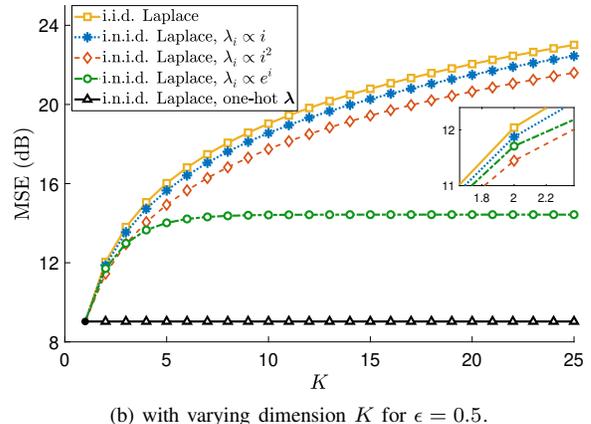}
			\caption{with varying dimension $K$ for $\epsilon=0.5\pspp$.}
			\label{fig:inid_lapl_K_vs_mse}
		\end{subfigure}
		\caption{Performance of i.i.d. and i.n.i.d. 
			$\epsilon$-DP 
			Laplace mechanisms under various sensitivity profiles with unit $\ell_1^{}$-sensitivity.}
		\label{fig:inid_lapl}
	\end{figure*}
	
	The performance of i.n.i.d. Gaussian mechanism with varying dimension $K$ is studied next. The MSE curves for different $K$ are given in  Figure \ref{fig:inid_gauss_K_vs_mse} for the privacy parameters $\epsilon=0.5$ and $\delta=10^{-6}_{}\nspp\pspp$. Figure \ref{fig:inid_gauss_gini} shows the Gini coefficients with varying $K$ for various  sensitivity profiles. From Figure \ref{fig:inid_gauss_K_vs_mse}, we can observe that the MSE of i.n.i.d. mechanism pertaining to quadratic profile is better than that corresponding to linear profile, which in turn offers lesser MSE than uniform profile (which coincides with the MSE of i.i.d. mechanism). The exponential profile results in lesser MSE than the quadratic one for $K\geq 3$; for $K=2\pspp$, the quadratic profile is better (please see the inset plot in Figure \ref{fig:inid_gauss_K_vs_mse}) because the quadratic profile is more spread out than the exponential one when $K=2\pspp$, which is evident from the larger Gini coefficient of the quadratic profile in Figure \ref{fig:inid_gauss_gini}. These results are in accordance with Theorem \ref{thm:improve_gau_inid} that the most dispersed $\mbf{{\lambda}}$ is associated with the least MSE. 
	
	It can also be observed that the reduction in MSE of the i.n.i.d. mechanism over i.i.d. one improves with $K$.
	However, for large $K$, the incremental reduction in MSE is smaller for the linear and quadratic profiles; for instance, both these profiles give only $0.02\psp\mathrm{dB}$ improvement for $K=25$ compared to $K=20\pspp$. However, the exponential profile provides a substantial reduction in MSE with increasing $K$ compared to the i.i.d. mechanism. This is because the MSE for the i.i.d. case increases linearly with $K$, $\norm{\mbf{\sigma}_0^{}}_2^2=\tfrac{K\Delta_2^2}{{\mu}_0^{2}}\nspp\pspp$, whereas the MSE curve for the exponential profile saturates for large $K$ at $21.477\psp\mathrm{dB}\pspp$. 

	\subsection{Laplace mechanism}
	The MSE curves of the i.n.i.d. Laplace mechanism that guarantees $\epsilon$-DP with varying $\epsilon$ are plotted in Figure \ref{fig:inid_lapl_eps_vs_mse}, and Figure \ref{fig:inid_lapl_K_vs_mse} shows the MSE with varying $K$. As with the Gaussian case, i.n.i.d. noise always provides improvement over the i.i.d. noise, and the reduction in MSE improves with the increase in the dispersion of $\mbf{{\lambda}}\pspp$. Notably, in Figure \ref{fig:inid_lapl_eps_vs_mse}, we can see that the i.n.i.d. Laplace noise reduces the MSE by $0.546\psp\mathrm{dB}\pspp$, $1.39\psp\mathrm{dB}\pspp$, and $7.609\psp\mathrm{dB}$ consistently over all $\epsilon\pspp$, for the linear, quadratic, and exponential sensitivity profiles, respectively. Figure \ref{fig:inid_lapl_K_vs_mse} also depicts a similar trend as that of our simulations for the Gaussian mechanism in Figure \ref{fig:inid_gauss_K_vs_mse}. The i.n.i.d. mechanism for the exponential profile offers lesser MSE than that pertaining to quadratic and linear profiles for larger $K\nspp\pspp$, and the reduction in MSE improves with $K$ since the MSE saturates at $14.243\psp\mathrm{dB}\pspp$, which is $5.4\psp\mathrm{dB}$ above the MSE for one-hot $\mbf{{\lambda}}\pspp$. 
	
	\begin{figure}[ht]
		\centering
		\includegraphics[width=0.84\linewidth]{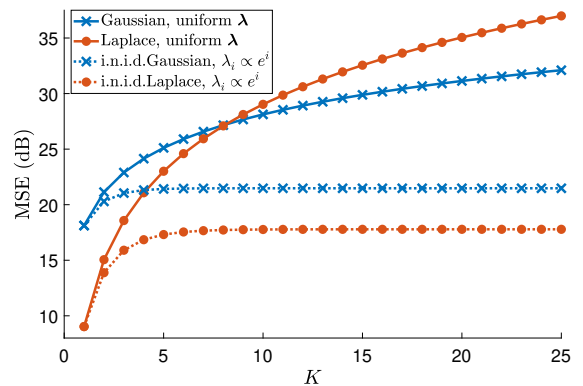}
		\caption{Comparison of $(\epsilon,\delta)$-DP Gaussian mechanism  and $\epsilon$-DP Laplace mechanism with varying dimension $K$ for $\epsilon=0.5$ and  $\delta=10^{-6}_{}$.}
		\label{fig:inid_lap_vs_gau}
	\end{figure}
	
	\subsection{Comparison of Gaussian and Laplace mechanisms}
	In Figure \ref{fig:inid_lap_vs_gau}, we compare the MSE pertaining to i.n.i.d. Laplace mechanism for $\epsilon=0.5$ with i.n.i.d. Gaussian  mechanism for $\epsilon=0.5$ and $\delta=10^{-6}_{}$ for different dimensions $K$. For this simulation, $\mbf{{\lambda}}$ is normalized to have $\Delta_2^{}=1\pspp$. Although the Gaussian mechanism is unable to provide the stronger $\epsilon$-DP guarantee with $\delta=0\pspp$, one of the reasons it is widely used is that it offers lesser MSE in higher dimensions than Laplace. This is the case for the uniform $\mbf{\lambda}$ (i.e., equal sensitivities), and we observe that the Laplace noise results in higher MSE than the Gaussian for $K\geq 9$ in Figure \ref{fig:inid_lap_vs_gau}. However, when ${\lambda}_i^{}\propto e^i_{}$, the i.n.i.d. Laplace mechanism offers lesser MSE than the Gaussian for all dimensions despite ensuring the stronger $\epsilon$-DP condition. Hence, the choice of noise distribution should not only be based on the dimension but also take into account the individual sensitivities. 
	
	\subsection{Comparison of staircase and Laplace mechanisms}
	We now compare the $\ell_1^{}$-errors (i.e., mean absolute errors, MAE) of the i.i.d. and i.n.i.d. Laplace mechanisms with that of the staircase mechanism under $\epsilon$-DP in two dimensions. The staircase density is piecewise flat with an exponential envelope, as shown in \cite[Figs. 1 and 2]{geng2015staircase}; since the density function decays exponentially in the tails, it can ensure pure DP \cite{tian2018selective}, like Laplace. Also, the corresponding privacy loss RV is discrete. In two dimensions, the staircase mechanism guarantees $\epsilon$-DP with the minimum $\ell_1^{}$-error \cite[Theorem 1]{geng2015staircase}; it adds \textit{correlated} noise but with only the knowledge of global $\ell_1^{}$-sensitivity. Accordingly, it outperforms the i.i.d. Laplace mechanism, which operates with just the knowledge of global $\ell_1^{}$-sensitivity. However, since the staircase mechanism does not account for the imbalance in the sensitivities, it results in more $\ell_1^{}$-error compared to the i.n.i.d. Laplace mechanism provided with the sensitivity profile. This is evident from the results presented in Table \ref{tab:staircase_and_lapl} for the sensitivity profile $\mbf{\lambda}=[0.85\,\ 0.15]^{\top}_{}$. Also, it can be observed that the gap is large in the high privacy regime. The results clearly show that whenever the coordinate-wise sensitivities are non-uniform, they have to be exploited. 
	\begin{table}[h!] 
		\caption{$\ell_1^{}$-errors of $\epsilon$-DP staircase and Laplace mechanisms with varying $\epsilon$ in $K=2$ dimensions for $\mbf{\lambda}=[0.85\,\ 0.15]^{\top}_{}$.}
		\label{tab:staircase_and_lapl}
		\setlength{\tabcolsep}{3.4pt}
		\begin{center}
			\begin{tabular}{?c?c|c|c|c|c|c?} 
				\clineB{1-7}{2}
				$\mbf{\epsilon}$	&	\textbf{0.5}	&	\textbf{1}	&	\textbf{1.5}	&	\textbf{2}	&	\textbf{2.5}	&	\textbf{3} \\[1ex] 
				\clineB{1-7}{2}
				\specialrule{1pt}{2pt}{0pt}	
				
				\textbf{Staircase}	&	3.9962	&	1.9862	&	1.3050	&	0.9546	&	0.7366	&	0.5856 \\[1ex]
				\clineB{1-7}{1}
				
				\textbf{i.i.d. Laplace}	&	4	&	2	&	1.3333	&	1	&	0.8	&	0.6667 \\[1ex]
				\clineB{1-7}{1}
				
				\textbf{i.n.i.d. Laplace}	&	3.4283	&	1.7141	&	1.1428	&	0.8571	&	0.6857	&	0.5714 \\[1ex] 
				\specialrule{1pt}{0pt}{0pt}
			\end{tabular}
			\vspace{-0.75ex}
		\end{center}
	\end{table}
	
	Note that the optimality of the staircase mechanism for $K>2$ is only a conjecture \cite{geng2015staircase}, and the $\ell_1^{}$-error is not characterized for $K>2\pspp$; one can add independent noise samples from the staircase density to each coordinate of the high-dimensional query output \cite{geng2015staircasearxiv}. The asymptotic analysis similar to that in Theorem \ref{thm:asymp} indicates that the i.n.i.d. Laplace mechanism offers $O({\nu})$ reduction over the $\ell_1^{}$-error of its i.i.d. counterpart\footnote{$\text{MAE}_{\,\text{i.n.i.d.}}^{\pspp(\mathcal{L})} = \frac{1}{\epsilon}\big\Vert{\mbf{{\lambda}}_{}^{\circ \nspp\frac{1}{2}}}\big\Vert_1^{2}	=\tfrac{K^2_{}\Delta}{\epsilon} \nspp\bigg(\!\tfrac{1}{K}\summ_{i=1}^K \!\sqrt{\frac{{\lambda}_i^{}}{\Delta}}\psp\bigg)^{\! 2}_{} \leq\tfrac{K^2_{}\Delta}{\epsilon}\times {\nu}	\pspp$.}; it also results in the same reduction over the $\ell_1^{}$-error of the staircase mechanism in the high privacy regime, as the performances of staircase and i.i.d. Laplace mechanisms match  as $\epsilon\to0$ \cite{geng2016optimalstaircase,geng2015staircase} (recall from our results for $K=2$ that the difference is substantial in this regime).
	
	\begin{figure*}[ht]
		\centering
		\begin{minipage}{0.5\linewidth}
			\begin{center}
				\centerline{\includegraphics[width=0.8\columnwidth]{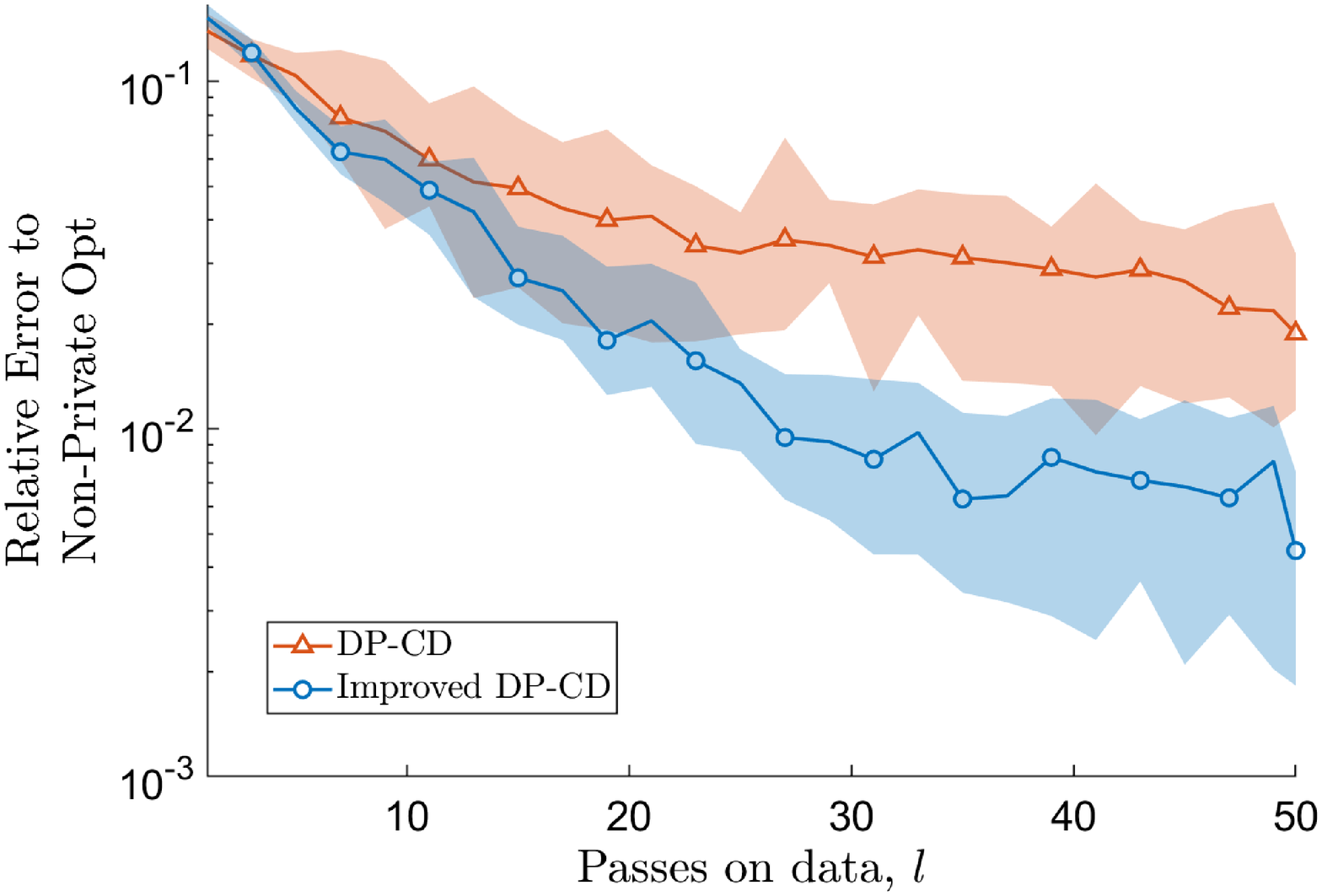}}
				\caption{Performance of i.n.i.d. noise in DP coordinate descent for $\ell_1^{}$-regularized linear regression on California dataset.}
				\label{fig:dpcd_calif_results}
			\end{center}
		\end{minipage}%
		\begin{minipage}{0.5\linewidth}
			\begin{center}
				\centerline{\includegraphics[width=0.8\columnwidth]{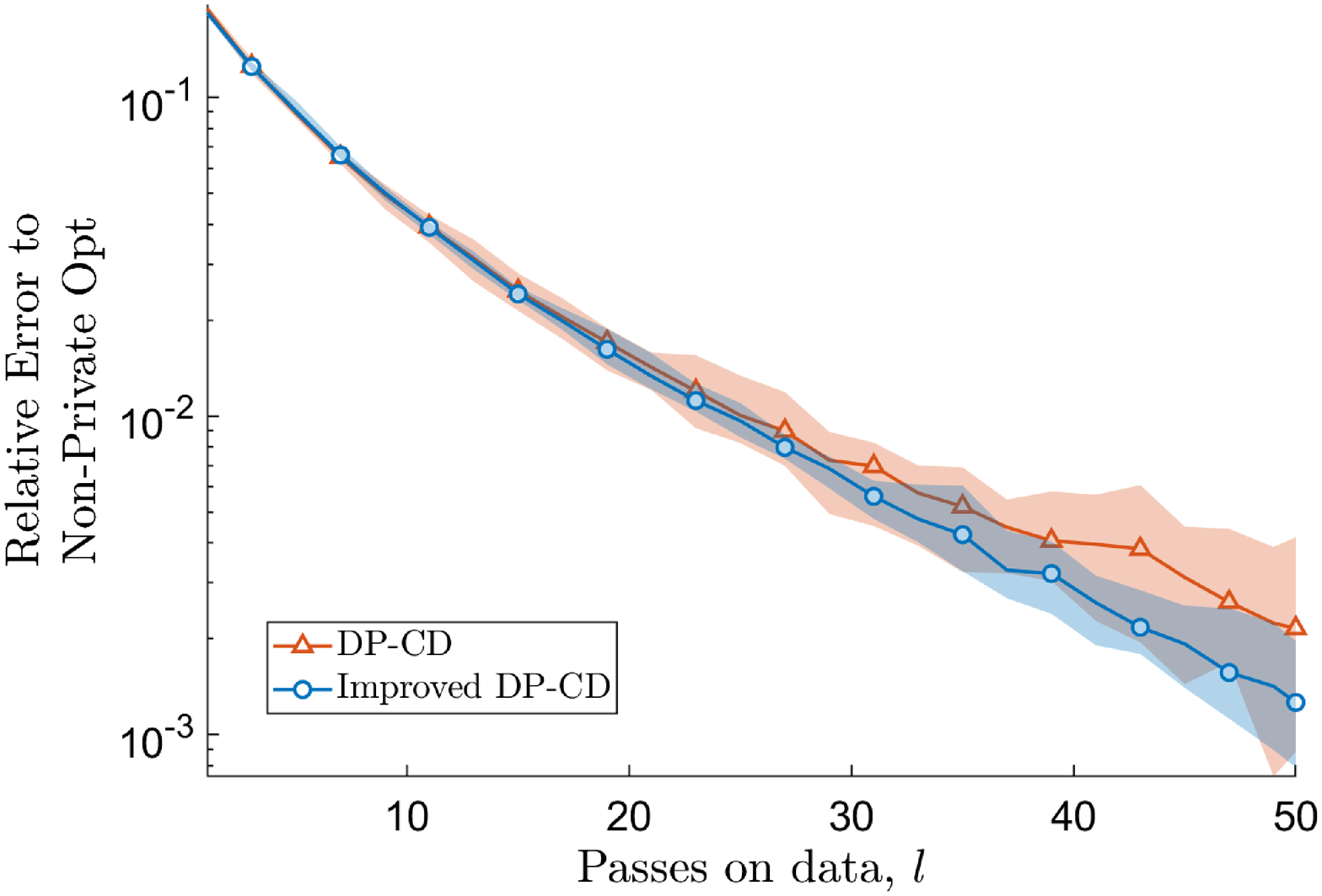}}
				\caption{Performance of i.n.i.d. noise in DP coordinate descent for  $\ell_2^{}$-regularized logistic regression on Electricity dataset.}
				\label{fig:dpcd_elec_results}
			\end{center}
		\end{minipage}
	\end{figure*}

	\section{Applications and discussions}\label{sec:appn}
	In this section, we demonstrate  the effectiveness of the i.n.i.d. noise in three private machine learning problems. 
	We also discuss the strengths and limitations of the proposed schemes.
	
	\subsection{DP coordinate descent (DP-CD)}\label{sec:appn_dpcd}
	We first consider empirical risk minimization through DP-CD \cite{mangold2022differentially}, where gradient updates are perturbed with noise to guarantee DP. Coordinate descent leverages the non-uniformity in coordinate-wise smoothness of the objective function to reduce the number of iterations by using different learning rates for each coordinate of the parameter; hence, it is an apt setting for validating i.n.i.d. mechanisms. 
	
	In \cite{mangold2022differentially}, the authors proposed adding non-identical noise to the \textit{clipped} coordinates of the gradient updates. However, the noise scales are set proportional to the clipping constants, and hence, it is an application of the SPR mechanism in Definition \ref{defn:spr_mechanism}. As discussed in Section \ref{sec:analysis}, the proposed i.n.i.d. noise addition exploits the coordinate-wise disparity better and is more appropriate than SPR mechanisms. We observe the same for the case of DP-CD; we label the application of the proposed i.n.i.d. scheme to the DP-CD as `improved DP-CD.' 
	
	We compare the performances of DP-CD and improved DP-CD under Gaussian noise\footnote{Additional information on the simulation setting is in Appendix \ref{appx:appn}.} for (i) linear regression with $\ell_1^{}$-regularization (i.e., LASSO) on California dataset \cite{californiadataset} and (ii) logistic regression with $\ell_2^{}$-regularization on Electricity dataset \cite{electricitydataset}. The benchmark for the comparison is the relative error to the (non-private) optimal objective value, i.e., $\big(\mathcal{J}(\mbf{\theta}^{(l)}_{\texttt{priv}}; \mathcal{D})-\mathcal{J}(\mbf{\theta}^*_{}; \mathcal{D})\big)/\mathcal{J}(\mbf{\theta}^*_{}; \mathcal{D})$, where $\mathcal{J}$ is the objective function of the ERM problem, $\mbf{\theta}^*_{}$ is the (non-private) optimal parameter, and $\mbf{\theta}^{(l)}_{\texttt{priv}}$ is the parameter estimate from the DP-CD after $l$ passes on the data. Figure \ref{fig:dpcd_calif_results} and Figure \ref{fig:dpcd_elec_results} show the relative errors as a function of the number of passes on the data. As in \cite{mangold2022differentially}, five random trials are performed for each combination of hyperparameters, and the lowest relative error averaged across the trials is plotted along with the error bars. From the results, it is evident that the proposed i.n.i.d. noise addition improves the performance of DP-CD. For instance, after fifty passes on the data, the average relative error of the DP-CD is $1.88\% $, whereas, for our improved DP-CD, it is only $0.45\%\pspp$; on the electricity dataset, the improved DP-CD offers a lower relative error of $0.13\%$ compared to $0.22\%$ of DP-CD.
	
	In Appendix \ref{appx:appn}, we present the results for the case where the sensitivity profile is privately determined from the dataset using a fraction of the privacy resource available. 
		
	\subsection{DP principal component analysis (DP-PCA)}\label{sec:appn_dppca}
	Estimating the subspace spanned by the users' data while preserving their privacy is a well-studied problem in literature \cite{dwork2014analyze,mangoubi2022re}. Let $\mbf{X}=[\,\mbf{x}_1^{}\ \mbf{x}_2^{} \cdots\ \mbf{x}_N^{}\,]\in\mathbb{R}^{M\times N}_{}$ be the dataset, where $\mbf{x}_i^{}\in\mathbb{R}^M_{}$, $i \in \mathbb{N}_N^{}$ is the data entry of the $i$-th user comprising of $M$ features, which we assume to be of unit $\ell_2^{}$-norm. The goal is to privately estimate the principal components $\mbf U\in \mathbb{R}^{M\times r}_{}$, where $r$ is the rank of the subspace. Without the privacy constraint, $\mbf U$ is obtained from the $r$ dominant eigenvectors of $\mbf R = \mbf{X}\mbf{X}^\top_{}$. Note that the entries of $\mbf{R}$ capture the relations between different features (i.e., rows of $\mbf{X}$). The direct way to ensure DP  is to perturb $\mbf R $ before estimating the eigenvectors. In \cite{dwork2014analyze}, the $\ell_2^{}$-sensitivity, when a user gets added or removed, is calculated as 1. However, due to the symmetry, we can treat $\mbf R $ as the query with upper triangular entries of $K=M(M+1)/2$ dimensions, add noise to them, and replicate the values to the lower triangle before releasing the result. Thus, the per-coordinate sensitivities of the diagonal and off-diagonal entries can respectively be set as $\frac{1}{\sqrt{K}}$ and $\frac{1}{\sqrt{2K}}$ since the off-diagonal entries appear twice. For i.n.i.d. mechanisms, we clip the entries of $\mbf{R}_i^{}=\mbf{x}_i^{}\mbf{x}_i^\top$ and add noise to upper triangular entries of $\widetilde{\mbf R}=\sum_{i=1}^N\operatorname{clip}\big(\mbf{R}_i^{}\big)$. 
	
	We gauge the performance in terms of Subspace Recovery Error (SRE) \cite{rahmani2017coherence}, $\text{SRE}=\tfrac{\lVert(\mathbf{I}-\hat{\mathbf{U}}\hat{\mathbf{U}}^{\top}_{})\mathbf{U}\rVert_{\text{F}}^{}}{\lVert\mathbf{U}\rVert_{\text{F}}^{}} \pspp$, where $\hat{\mathbf{U}}$ is the orthogonal basis of the private subspace estimated by the algorithm. We report the average of these metrics over 1000 trials. We consider the synthetic data from the fully random model\cite{soltanolkotabi2012geometric} with parameters $N=100\pspp$, $N=1000\pspp$,  and $r=10\pspp$. The results for the setting  $\delta=1/N^2_{}=10^{-6}_{}$ for Gaussian and $\delta=0$ for Laplace are shown in Table \ref{tab:dp_pca} for three different values of $\epsilon$. The i.n.i.d. Gaussian mechanism performs the best in terms of SRE; the i.n.i.d. Laplace comes close to that of i.i.d. Gaussian, but it ensures the strong $\epsilon$-DP unlike the latter. Thus, the distortion of the relations between features is imminent due to the privacy requirements, but the results indicate that such deviations can be reduced with the use of non-identical noise.
	
	\begin{table}
		\begin{center}
			\caption{DP-PCA performance of i.i.d. and i.n.i.d. mechanisms on various datasets.}
			\label{tab:dp_pca}
			\setlength{\tabcolsep}{4pt}
			\begin{tabular}{? C{4em} | C{3em} ? C{5em} | C{5em} | C{5em} ?}
				\clineB{3-5}{2}
				\multicolumn{1}{c}{} &\multicolumn{1}{c?}{} & $\mbf{\epsilon=1}$  & $\mbf{\epsilon=2}$ & $\mbf{\epsilon=5}$  \\
				\clineB{3-5}{2}
				\specialrule{1pt}{2pt}{0pt}
				
				\multirow{2}{*}{\textbf{Gaussian}}
				& i.i.d. &  0.4762 &0.2484   & 0.1067 \\
				\cline{2-5}
				
				& i.n.i.d. & \textbf{0.3269} &\textbf{0.1728}   &  \textbf{0.0752} \\
				\clineB{1-5}{2}
				\specialrule{1pt}{2pt}{0pt}	
				
				\multirow{2}{*}{\textbf{Laplace}}
				& i.i.d. & 0.6185 &0.3272   & 0.1389 \\
				\cline{2-5}
				
				& i.n.i.d. & 0.5034 &0.2523   & 0.1093 \\
				\clineB{1-5}{2}
				\specialrule{1pt}{0pt}{1pt}
			\end{tabular}
		\end{center}
	\vspace{-4ex}
	\end{table}
	
	\subsection{Deep learning with DP through group-wise clipping}
	In this application, we consider the ERM problem of training deep learning models with DP guarantees by adopting DP stochastic gradient descent (DP-SGD) \cite{abadi2016deep,bassily2014private}. Similar to DP-CD, in each iteration, the gradient updates are clipped to limit the sensitivity, and the noise is added to the average of the clipped gradients. The gradients are computed over the Poisson subsampled dataset, and we account for the privacy gains from such subsampling via R{\'e}nyi DP \cite{zhu2019poission, mironov2019renyi}. 
	
	Recently, a lot of work has focussed on performing group-wise clipping where the gradient coordinates corresponding to parameters of a layer are treated as a group \cite{he2022exploring,mcmahan2018learning,mcmahan2018general}. This substantially reduces the computations and improves the speed compared to global clipping. However, these mechanisms set the noise scale for a layer only based on the clipping levels of that layer, and hence, are the instances of SPR mechanisms. We now provide the optimal scale parameters based on our i.n.i.d. framework. For the $M$-layer neural network, let $C_{m}^{(p)}$ denote the clipping budget (in $\ell_p^{}$-norm) available to the $m$-th layer with $K_m^{}$ parameters, $m\in\mathbb{N}_M^{}$. We clip the gradient of the $i$-th parameter in the $m$-th layer to have a maximum absolute value of $\lambda_{i}^{(m)}=\frac{C_m^{(p)}}{K_m^{1/p}}$; thus, the  maximum change brought forth by the addition/deletion of a data point in the $\ell_p^{}$-norm of the $m$-th layer's gradient is $C_m^{(p)}$. Note that, for the network where $K_m^{}$ is different for different layers, the sensitivity profile $\big\{\lambda_i^{(m)}\,\big|\, i\in K_m^{},\, m\in\mathbb{N}_m^{}\big\}$ is disparate; we split the available privacy resources equally for all the gradient updates, and this disparate sensitivity profile results in unequal allocation of each update's resource to the coordinates. We note that a later work \cite{xiao2023geometry} has also given an identical result for the Gaussian case, where they also consider learning projectors from publicly available datasets, but our results hold for a general setting without the need for public data and provide the scales for Laplace mechanism as well.
	
	We demonstrate the performance results for the classification problem with ResNet-20 on the CIFAR-10 dataset. We consider flat per-layer clipping, i.e., $C_{m}^{(p)}$ are set equal so that their $\ell_p^{}$-norm is equal to the  overall clipping budget $C_0^{(p)}=1\pspp$. Note that even in such a case, the noise parameters will not be identical as the number of parameters in each layer varies. We consider the subsampling ratio of $0.02$ and train the algorithm for 100 epochs. We compare the test accuracy with that of the global clipping \cite{abadi2016deep} with the clipping threshold $C_0^{(p)}=1$ and the SPR equivalent of the per-layer flat clipping \cite{he2022exploring}. The results are presented in Table \ref{tab:dp_dl}.
	\begin{table}
		\begin{center}
			\caption{Test accuracy for deep learning with i.i.d. and i.n.i.d. mechanisms.}
			\label{tab:dp_dl}
			\setlength{\tabcolsep}{4pt}
			\begin{tabular}{? C{7em} | C{3.25em} ? C{5em} | C{5em} | C{5em} ?}
				\clineB{3-5}{2}
				\multicolumn{1}{c}{} &\multicolumn{1}{c?}{} & 
				\begin{tabular}{c}
					\hspace{-0.35em}\textbf{i.i.d.}
					\\[-0.75em]
					\hspace{-0.35em}\textbf{(Global)}
				\end{tabular}  & 
				\begin{tabular}{c}
					\hspace{-0.35em}\textbf{SPR i.n.i.d.}
					\\[-0.75em]
					\hspace{-0.35em}\textbf{(per-layer)}
				\end{tabular} &  
				\begin{tabular}{c}
					\hspace{-0.35em}\textbf{Our i.n.i.d.}
					\\[-0.75em]
					\hspace{-0.35em}\textbf{(per-layer)}
				\end{tabular}  \\
				\clineB{3-5}{2}
				\specialrule{1pt}{2pt}{0pt}
				\multirow{2}{*}{\begin{tabular}{c}
						\\[-2.5em]
						\textbf{Gaussian}
						\\[-0.75em]
						($\delta=10^{-5}$)
				\end{tabular}}
				& {${\epsilon=5}$} & 63.26 & 42.90 &  57.86 \\
				\cline{2-5}
				& {${\epsilon=8}$} & 68.55 & 47.16 & 61.58 \\
				\clineB{1-5}{2}
				\specialrule{1pt}{2pt}{0pt}	
				\multirow{2}{*}{\begin{tabular}{c}
						\\[-2.5em]
						\hspace{-0.25em}\textbf{Gaussian}
						\\[-0.75em]
						\hspace{-0.25em}($\delta=4\cdot10^{-10}$)
				\end{tabular}}
				& {${\epsilon=5}$} & 59.03 & 42.27 &  50.33 \\
				\cline{2-5}
				& {${\epsilon=8}$} & 60.47 & 44.02  &  57.65 \\
				\clineB{1-5}{2}
				\specialrule{1pt}{2pt}{0pt}	
				\multirow{2}{*}{\textbf{Laplace}}
				& {${\epsilon=5}$}  & 62.17 & 42.05 & 59.02 \\
				\cline{2-5}
				& {${\epsilon=8}$} & 67.46 & 46.08 & 63.46 \\
				\clineB{1-5}{2}
			\end{tabular}
		\end{center}
	\vspace{-3.7ex}
	\end{table}
	
	From the results, it can be observed that the i.n.i.d. mechanisms with the proposed set of noise parameters perform better than the SPR counterparts, but there is a deterioration  compared to the performance of global clipping, as expected with per-layer clipping.  It can be observed that the proposed i.n.i.d. Laplace mechanism results in a better accuracy compared to the corresponding Gaussian mechanism; in particular, it improves the accuracy by $1.16\%$ and $1.88\%$, respectively for $\epsilon=5$ and $\epsilon=8\pspp$. However, for the SPR and i.i.d. equivalents, the Laplace noise deteriorates the accuracy. These results corroborate our claim that the Laplace noise, with the right setting of parameters, can outperform Gaussian and guarantee the stronger $\epsilon$-DP.
	
	We also note that, in deep learning with DP, the `cryptographically small' requirement on $\delta$ \cite{dwork2014algorithmic} is discounted to get better accuracy. For the given dataset: $\delta=10^{-5}_{}$ is not cryptographically small; for $\delta=1/N^2=4\times 10^{-10}_{}$, i.i.d. Gaussian noise with global clipping results in only $60.47\%$ accuracy when $\epsilon=8$, which is lesser than the $63.46\%$ accuracy obtained using the proposed i.n.i.d. Laplace mechanism with per-layer clipping guaranteeing stronger $(8,0)$-DP. This further highlights the significance of Laplace noise in this application.
	
	\subsection{Discussions}
	
	Throughout the article, we have observed that the non-identical noise addition provides better utility than the i.i.d. mechanisms in a wide range of settings and applications. The key advantage of the i.n.i.d. noise addition is that it renders utility efficient $\epsilon$-DP in higher dimensions feasible since Laplace can outperform Gaussian even in higher dimensions. Since $\delta$ needs to be set cryptographically smaller in the size of the dataset for $(\epsilon, \delta)$-DP \cite{dwork2014algorithmic}, this result is powerful as it renders utility efficient privacy preservation on large databases through i.n.i.d. Laplace mechanism. Also, the theoretical results presented in Section \ref{sec:mse_redn} motivate designing queries that result in useful yet meaningful sensitivity profiles to improve utility. For instance, in the DP-PCA application in Section \ref{sec:appn_dppca}, we have restructured the query to yield a disparate sensitivity profile	by harnessing the symmetry of the matrix to be perturbed.
	
	However, though the sensitivity profile renders formal analysis of non-identical noise mechanisms possible, it might be difficult to determine it in practice: For the deep learning application, the sensitivity profile is heuristically set through an ad-hoc clipping budget allocation, as is customary in existing works, and for coordinate descent, one may have to spend a fraction of the available privacy resources to estimate $\mbf{\lambda}$ (please see the supplementary material). We hope that this article motivates future works to study this problem in detail. Also, when the coordinates are strongly correlated, the approximation $\Delta_p^{}=\norm{\mbf{\lambda}}_p^{}$ would fail, and one cannot guarantee that the proposed i.n.i.d. schemes perform better than the i.i.d. counterparts; practitioners need to validate this assumption to show provable improvement.
	
	\section{Conclusions and future directions}\label{sec:conc}
	
	We have introduced i.n.i.d. noise addition to perturb the query results on databases to guarantee privacy. In particular, Gaussian and Laplace i.n.i.d. mechanisms are studied in detail. The use of i.n.i.d. noise offers more degrees of freedom with one scale parameter per coordinate, and the MSE can be minimized by exploiting the disparity in the sensitivities across the coordinates. The appropriate choices for the scale parameters for the i.n.i.d. Gaussian and Laplace mechanisms that result in the least perturbation for the required privacy guarantees have been derived. It has been shown theoretically and empirically that this choice of parameters improves the utility over the i.i.d. noise for a wide range of scenarios. We have also observed that the Laplace mechanism can result in lesser perturbation than Gaussian even in higher dimensions when i.n.i.d. noise is added. We have also shown that the Laplace mechanism can beat the staircase mechanism by exploiting coordinate-wise disparity. We further demonstrated that the proposed i.n.i.d. noise addition offers better performance in (a)  private empirical risk minimization through coordinate descent, (b) private principal component analysis, and (c) deep learning with differential privacy and per-layer clipping, and emphasized the utility of Laplace noise in deep learning.
	
	We hope that this work serves as a	starting point for other future works. The extension of ideas investigated in this article to non-numeric queries that do not fall under the framework of additive noise mechanism, for example, exponential mechanism and randomized response, is an interesting aspect to explore. Another interesting direction for future research is the use of different noise types across the coordinates to improve the utility; the right choice of unifying privacy framework  \cite{bun2016concentrated,mironov2017renyi,zhu2022optimal,dong2022gaussian,sommer2019privacy,koskela2020computing} that enables tight and easy analysis with heterogeneous noise types would be the key to this problem.
	
	\appendices
	\section{Deferred Proofs}\label{appx:proof_sec_inid_mech} 
	
	We first provide a result that is useful throughout our analysis.
	\begin{lemma}\label{lem:gau_priv_prof_monotonic}
		The function ${\varphi}_{\epsilon}^{}:\mathbb{R}_{++}^{}\to\mathbb{R}\pspp$, defined by
		\begin{equation} \label{eq:gau_priv_prof_fn}
			{\varphi}_{\epsilon}^{}({a})={Q}\!\left(\nsp\tfrac{\epsilon}{{a}}-\tfrac{{a}}{2}\nsp\right)\! -	e^{\epsilon}_{} {Q}\!\left(\nsp\tfrac{\epsilon}{{a}}+\tfrac{{a}}{2}\nsp\right)\!
			\psp, \end{equation}
		is a monotonic increasing function for any $\epsilon \geq 0 \pspp$.
	\end{lemma}
	
	\begin{proof}
		The lemma is proved by showing that $\tfrac{\mathrm{d}{\varphi}_{\epsilon}^{}}{\mathrm{d}{a}}\geq0\,\ \forall \,{a}>0 \nspp\pspp$. Using the Leibniz integral rule, we have $\tfrac{\mathrm{d}}{\mathrm{d}{b}}{Q}({b})=-\tfrac{e^{-{b}^2_{}/2}_{}}{\sqrt{2\pi}}\nspp\pspp$. Hence,
		\begin{align*}
			\tfrac{\mathrm{d}}{\mathrm{d}{a}}{\varphi}_{\epsilon}^{}({a})
			&=\tfrac{\mathrm{d}}{\mathrm{d}{a}}\!\left[{Q}\!\left(\nsp\tfrac{\epsilon}{{a}}-\tfrac{{a}}{2}\nsp\right)\! -	e^{\epsilon}_{} {Q}\!\left(\nsp\tfrac{\epsilon}{{a}}+\tfrac{{a}}{2}\nsp\right)\!\psp\right]
			\\&
			=\tfrac{1}{\sqrt{2\pi}}\exp\!\left(\nspp -\tfrac{1}{2}\big(\tfrac{\epsilon}{{a}}-\tfrac{{a}}{2}\big)^{\nsp 2}_{}\nspp\right)\!\nsp\left[\tfrac{\epsilon}{{a}^2_{}}+\tfrac{1}{2}\right]
			\\&\qquad
			-\tfrac{1}{\sqrt{2\pi}}\exp\!\left(\nspp \epsilon -\tfrac{1}{2}\big(\tfrac{\epsilon}{{a}}+\tfrac{{a}}{2}\big)^{\nsp 2}_{}\nspp\right)\!\nsp\left[\tfrac{\epsilon}{{a}^2_{}}-\tfrac{1}{2}\right]
			\\&
			=\tfrac{1}{\sqrt{2\pi}}\exp\!\left(\nspp -\tfrac{1}{2}\big(\tfrac{\epsilon}{{a}}-\tfrac{{a}}{2}\big)^{\nsp 2}_{}\nspp\right)\! \psp\geq 0\,\ \ \forall \,\,{a}>0 \nspp
			\psp. \qedhere
		\end{align*}
	\end{proof}
	
	\begin{proof}[Proof of Lemma \ref{lem:priv_gau_inid}]
		The equivalent privacy loss function for the i.n.i.d. Gaussian mechanism is given as ${\zeta}_{\pspp \mbf{d}}^{}\nsp(\mbf{t})=\sum_{i=1}^K {\zeta}_{\pspp d_i^{}}^{}\nsp(t_i^{})\pspp$, where ${\zeta}_{\pspp d_i^{}}^{}\nsp(t_i^{})=\log\frac{{g}_{\pspp T_i}^{}(t_i^{})}{{g}_{\pspp T_i}^{}(t_i^{}+d_i^{})}\nspp\pspp$. Since the noise density is given as ${g}_{\pspp T_i}^{}(t_i)= \frac{1}{\sqrt{2\pi\sigma_i^2}}\exp\!\left(\nspp-\tfrac{t_i^2}{2\sigma_i^2}\nspp\right)\!\pspp\pspp$, we can deduce that ${\zeta}_{\pspp \mbf{d}}^{}\nsp(\mbf{t})=\summ_{i=1}^K \tfrac{t_i^{}d_i^{}}{\sigma_i^2}+ \tfrac{d_i^2}{2\sigma_i^2} \pspp$. We know that $\sum_{i=1}^K \tfrac{T_i^{}d_i^{}}{\sigma_i^2} \sim \mathcal{N}\!\left(\nspp 0, \|{\mbf{m}}\|_2^2 \right)\! \psp\pspp$, where $\mbf{m}=\diag(\mbf{\sigma})^{-1}_{}\mbf{d}\pspp$. Therefore, the privacy loss random variable ${\zeta}_{\pspp \mbf{d}}^{}\nsp(\mbf{T})$ is also Gaussian, and hence, $\mathbb{P}\{{\zeta}_{\pspp \mbf{d}}^{}\nsp(\mbf{T})\geq\epsilon\}={Q}\!\left(\nsp\tfrac{\epsilon}{\|{\mbf{m}}\|_2^{}} \nsp -\nsp \tfrac{\|{\mbf{m}}\|_2^{}}{2}\nsp\right)\!$ and $\mathbb{P}\{{\zeta}_{\pspp -\mbf{d}}^{}\nsp_{}(\mbf{T})\leq-\epsilon\} = {Q}\!\left(\nsp\tfrac{\epsilon}{\|{\mbf{m}}\|_2^{}}+\tfrac{\|{\mbf{m}}\|_2^{}}{2}\nsp\right)\!\psp$. Hence, using \eqref{eq:balle_K}, the necessary and sufficient condition for $(\epsilon,\delta)$-DP is 
		\begin{equation}\label{eq:inid_gau_priv_prof0}	{Q}\!\left(\nsp\tfrac{\epsilon}{\|{\mbf{m}}\|_2^{}}-\tfrac{\|{\mbf{m}}\|_2^{}}{2}\nsp\right)\! -e^{\epsilon}_{} {Q}\!\left(\nsp\tfrac{\epsilon}{\|{\mbf{m}}\|_2^{}}+\tfrac{\|{\mbf{m}}\|_2^{}}{2}\nsp\right)\! \leq \delta \psp,
		\end{equation}
		which must hold for every pair of neighbouring datasets. From Lemma \ref{lem:gau_priv_prof_monotonic}, we know that the function at the left is a monotonic increasing function in $\|{\mbf{m}}\|_2^{}\pspp$, which in turn is a monotonic increasing function in each of $|d_i^{}|\pspp$. Also, $\supoverneighb \norm{\mbf{{m}}}_2^{} \sqrt{\sum_{i=1}^K \tfrac{{\lambda}_i^2}{\sigma_i^2}} \triangleq {\mu}  \pspp$. Thus, by taking the supremum of \eqref{eq:inid_gau_priv_prof0} over every pair of neighbouring datasets $\neighbdsets$, and using the bound $\mu$, we obtain the condition \eqref{eq:gauss_priv_cond}.
	\end{proof}
	
	\begin{proof}[Proof of Corollary \ref{thm:priv_lap_inid_suff}]
		We start with the following necessary and sufficient condition for the additive noise mechanism to guarantee $(\epsilon, \delta)$-DP \cite[Theorem 5]{balle2018improving}:
		\begin{equation*}
			\int_{\mathbb{R}^K_{}} \!\big[{g}_{\pspp\mbf{T}}^{}\nspp(\mbf{t})-e^{\epsilon}_{}{g}_{\pspp\mbf{T}}^{}\nspp(\mbf{t}\nspp+\nspp \mbf{d})\big]_{\nsp+}^{} \pspp \mathrm{d}\mbf{t}  \leq \delta 	\ \,\,\ \forall \ \mbf{d} \in \{\mbf{a}\,\big|\pspp\abs{a_i^{}}\leq \lambda_i^{}\}\psp,
		\end{equation*}
		where ${g}_{\pspp\mbf{T}}^{}\nspp(\mbf{t})=2^{-K}_{}\big(\nsp\prod_{i=1}^{K}\beta_i^{}\big)^{-1}_{} \exp\big(\!-\nspp\norm{\mbf{t}\oslash\mbf{\beta}}_1^{}\nsp\big)\nsp$ is the noise density. Due to triangle inequality, we have
		\begin{equation*}
			\norm{(\mbf{t}\nsp+\nsp \mbf{d})\nsp\oslash\nsp\mbf{\beta}}_1^{}\nspp 
			\nspp\leq\nspp \norm{\mbf{t}\nsp\oslash\nsp\mbf{\beta}}_1^{}\nspp \nsp+\nsp \norm{\mbf{d}\nsp\oslash\nsp\mbf{\beta}}_1^{}\nspp
			\nspp\leq\nspp \norm{\mbf{t}\nsp\oslash\nsp\mbf{\beta}}_1^{}\nspp \nsp+\nsp \norm{\mbf{\lambda}\nsp\oslash\nsp\mbf{\beta}}_1^{}\nspp;
		\end{equation*}
		thus, $ \exp\big(\!-\nspp\norm{(\mbf{t}\nsp+\nsp \mbf{d})\nsp\oslash\nsp\mbf{\beta}}_1^{}\nsp\big)\nsp \geq \exp\big(\!-\nspp\norm{\mbf{t}\nsp\oslash\nsp\mbf{\beta}}_1^{}\nspp \nsp-\nsp \norm{\mbf{\lambda}\nsp\oslash\nsp\mbf{\beta}}_1^{}\nsp\big)\nsp\pspp$. Using this, we obtain the sufficient condition for $(\epsilon,\delta)$-DP as
		\begin{equation*}
			\begin{split}
				\int_{\mathbb{R}^K_{}}\nsp\!\big[\exp\big(\nsp-\nspp\norm{\mbf{t}\nsp\oslash\nsp\mbf{\beta}}_1^{}\nsp\big) -\exp\big(\epsilon \nsp-\nsp\nspp\norm{\mbf{t}\nsp\oslash\nsp\mbf{\beta}}_1^{}\nspp 
				&\nsp-\nsp \norm{\mbf{\lambda}\nsp\oslash\nsp\mbf{\beta}}_1^{}\nsp\big)\nspp\big]_{\nsp+}^{} \pspp \mathrm{d}\mbf{t}  
				\\ &
				\hspace{-2ex}
				\leq 
				\delta \times 2^{K}_{}\textstyle\prod_{i=1}^{K}\beta_i^{}
				\pspp.
			\end{split}
		\end{equation*}
		We observe that the integrand is zero when $\norm{\mbf{\lambda}\nsp\oslash\nsp\mbf{\beta}}_1^{} \leq \epsilon\pspp$, and we will get the condition for $\epsilon$-DP. But when $\norm{\mbf{\lambda}\nsp\oslash\nsp\mbf{\beta}}_1^{} \leq \epsilon\pspp$, the $[\psp\cdot\psp]_+^{}$ operator can be dropped from the integrand; with some simplifications, we get the sufficient condition $1-\exp\big(\epsilon \nsp-\nsp\nspp\norm{\mbf{\lambda}\nsp\oslash\nsp\mbf{\beta}}_1^{}\nsp\big)\nspp\leq \delta$, which can be re-expressed   as 
		$\sum_{i=1}^K \tfrac{{\lambda}_i^{}}{{\beta}_i^{}}=\epsilon-\log(1-\delta)\pspp$. Proceeding similar to the steps following \eqref{eq:boundary_lap}, we get the result.
	\end{proof}
	
	\section{Solving the optimization problem (P2)}\label{appx:bisection_method}
	
	To solve the problem (P2) efficiently, let us consider the function
	\begin{equation*}\label{eq:gau_priv_prof_fn_modif}
		\mathcal{B}_{\epsilon,\delta}^{}({\mu})\nspp= \nspp
		{\varphi}_{\epsilon}^{}({\mu})-\delta=\nspp {Q}\!\left(\nsp{\tfrac{\epsilon}{{\mu}}}\nsp-\nsp\tfrac{{\mu}}{2}\nsp\right)\!\pspp -	e^{\epsilon}_{} {Q}\!\left(\nsp{\tfrac{\epsilon}{{\mu}}}\nsp+\nsp\tfrac{{\mu}}{2}\nsp\right)\! -\delta  \pspp.
	\end{equation*}
	From Lemma \ref{lem:gau_priv_prof_monotonic}, we know that $\mathcal{B}_{\epsilon,\delta}^{}$ is a monotonic increasing function on $\mathbb{R}_{++}^{}\pspp$. Since ${\mu}_0^{}$, which is the solution to the problem (P2), corresponds to the largest ${\mu}$ so that ${\varphi}_{\epsilon}^{}({\mu}) \leq \delta\pspp$, ${\mu}_0^{}$ is the (positive) root of $\mathcal{B}_{\epsilon,\delta}^{}\pspp$, i.e., $\mathcal{B}_{\epsilon,\delta}^{}({\mu}_0^{}) =0 $. We focus on obtaining this root. 
	
	Since $\mathcal{B}_{\epsilon,\delta}^{}$ is monotonic increasing, it is also quasi-convex, and the root ${\mu}_0^{}$ can be obtained using the bisection method \cite{burden2011numerical,boyd2004convex}. Bisection method is iterative. It begins with an interval in which the function $\mathcal{B}_{\epsilon,\delta}^{}$ changes its sign, and in each iteration, it shrinks the interval to half its current length so that the function still changes the sign in the new interval. The procedure can be terminated once the length of the interval gets smaller than the required level of accuracy in the root.
	
	We know that $\mathcal{B}_{\epsilon,\delta}^{}({\mu})$ is bounded above and below by $\mathcal{C}_{\epsilon,\delta}^{}({\mu})$ and $\mathcal{C}_{\epsilon,\delta'_{}}^{}({\mu})$, respectively, where $\mathcal{C}_{\epsilon,{a}}^{}({\mu})= {Q}\!\left(\nsp{\tfrac{\epsilon}{{\mu}}}\nsp-\nsp\tfrac{{\mu}}{2}\nsp\right)\!\pspp -	{a}\pspp$, and  $\delta'_{}=\delta +	e^{\epsilon}_{}Q(\nspp\sqrt{2\epsilon}\psp)\pspp$. Note that $\mathcal{C}_{\epsilon,\delta}^{}$ is also a monotonic increasing function. Thus, the (positive) root of $\mathcal{C}_{\epsilon,\delta}^{}$, given by ${\mu}^{\nspp({l})}_0=\mathcal{R}_{\epsilon}^{}(\delta)\pspp$, lower bounds  ${\mu}_0^{}\pspp$, where $\mathcal{R}_{\epsilon}^{}(\delta)=\sqrt{(Q^{-1}\nspp(\delta))^2_{}+2\epsilon}- Q^{-1}\nspp(\delta)\pspp$; likewise, ${\mu}^{\nspp({\nspp u\nspp})}_0=\mathcal{R}_{\epsilon}^{}(\delta'_{})\pspp$, serves as an upper bound for ${\mu}_0^{}$. Hence, $\mathcal{B}_{\epsilon,\delta}^{}({\mu})$ changes its sign over ${\mu}\in\big[{\mu}^{\nspp({l})}_{0},{\mu}^{\nspp({\nspp u\nspp})}_{0}\big]\pspp$.  We can choose this interval as the initial interval for the bisection method for obtaining the root ${\mu}_0^{}\pspp$. Once the interval gets shorter than the tolerance level in the bisection method, we take ${\mu}_0^{}$ as the lower limit of that interval since it holds that ${\varphi}_{\epsilon}^{}\nspp\big({\mu}^{({l})}_{}\big)\nspp \leq \delta\pspp$. The procedure is outlined in Algorithm \ref{alg:bisection_method}. It converges linearly and finds ${\mu}_0^{}$ in $\log_{\pspp2}^{}\nspp\!\left(\nsp\tfrac{{\mu}^{\nspp({\nspp u\nspp})}_{0}-{\mu}^{\nspp({l})}_{0}}{{\mu}_{\texttt{tol}}}^{}\nsp\right)\!$ iterations, where  ${\mu}_{\texttt{tol}}^{}$ is the required level of accuracy in the estimate of ${\mu}_0^{}$\cite{burden2011numerical}.
	
	\begin{algorithm}[h!]
		\caption{Bisection method to solve (P2).}
		\label{alg:bisection_method}
		\begin{algorithmic}[1]
			\State {\bfseries Input:} privacy parameters $\epsilon\geq 0$ and $\delta\in[0,1]\pspp$, tolerance level ${\mu}_{\texttt{tol}}^{}\nspp\pspp$.
			\vspace{0.3em}
			\State ${\mu}^{({l})}_{}\psp\gets\sqrt{(Q^{-1}\nspp(\delta))^2_{}+2\epsilon}\nsp-\nsp Q^{-1}\nspp(\delta)\pspp$.
			\vspace{0.3em}
			\State ${\mu}^{({\nspp u\nspp})}_{}\gets\nspp\sqrt{(Q^{-1}\nspp(\delta'_{}))^2_{}\nsp+\nsp2\epsilon}\nsp-\nsp Q^{-1}\nspp(\delta'_{})\pspp$,  $\delta'_{}=\delta +	e^{\epsilon}_{}Q(\nspp\sqrt{2\epsilon}\psp)\pspp$. 
			\Repeat
			\State ${\mu}^{(\nspp {m}\nspp )}_{}\gets\big({\mu}^{({l})}_{}+{\mu}^{(\nspp {u}\nspp )}_{}\big)\big/\pspp2\pspp$.
			\If{ $\mathcal{B}_{\epsilon,\delta}^{}\big({\mu}^{(\nspp {m}\nspp )}_{}\big)>0$}
			\vspace{0.3em}
			\State ${\mu}^{(\nspp {u}\nspp )}_{}\gets {\mu}^{(\nspp {m}\nspp )}_{}$.
			\Else
			\State ${\mu}^{({l})}_{}\gets {\mu}^{(\nspp {m}\nspp )}_{}$.
			\EndIf
			\Until{${\mu}^{(\nspp {u}\nspp )}_{}-{\mu}^{({l})}_{}\leq {\mu}_{\texttt{tol}}^{}\nspp\pspp$.}
			\State {\bfseries Output:} ${\mu}_{0}^{} = {\mu}^{({l})}_{}\nsp\pspp$.
		\end{algorithmic}
	\end{algorithm}
	
	\section{Additional details and results on DP coordinate descent  (DP-CD)}\label{appx:appn}	
	We consider the empirical risk minimization problem, 
	\begin{equation*}\label{eq:erm}
		\underset{\mbf{{\theta}}\in\mathbb{R}^K_{}}\min\ \mathcal{J}(\mbf\theta; \mathcal{D}) \triangleq  \frac{1}{n}\nspp\summ_{n=1}^N J(\mbf{{\theta}}; \mathcal{D}_n^{})+ {\psi}(\mbf{{\theta}}) \psp,
	\end{equation*}
	where $\mbf{{\theta}}\in\mathbb{R}^K_{}$ is the model parameter to be optimized, $\mathcal{D}=(\mathcal{D}_1^{},\,\mathcal{D}_2^{},\,\ldots,\,\mathcal{D}_N^{})\in\mathcal{X}$ is the dataset of $N$ samples, and $\mathcal{D}_n^{}=(\mbf{x}_n^{},y_n^{})$ is the tuple of $n$-th user's attribute and label. Let $J:\mathbb{R}^K_{} \times \mathcal{X} \to \mathbb{R}$ be a convex and smooth loss function, and ${\psi}:\mathbb{R}^K_{} \to \mathbb{R}$ be a convex and separable regularizing function, ${\psi}(\mbf{{\theta}})=\sum_{i=1}^{K}{\psi}_i^{}(\theta_i^{})$. It is assumed that  the coordinate-wise smoothness constants of the objective function, $\{{M}_i^{}\}_{i=1}^{K}$, are available (for generalized linear models, we can obtain them from the data \cite[Sec. 5.2]{mangold2022differentially}; for the results corresponding to the case where these are unknown and have to be estimated privately, please see the supplementary material). The proximal operators corresponding to the regularizers are 
	\begin{equation*}
		\operatorname{prox}_{{\tau}_i^{}{\psi}_i^{}}^{} \nsp (\mbf{{\theta}})  = \underset{\mbf{w}\in\mathbb{R}^K_{}}{\operatorname{argmin}} \ \tfrac{1}{2}\norm{\mbf{w} - \mbf{{\theta}}}_2^2 + {\tau}_i^{}{\psi}_i^{}(w_i^{}) \psp, 
	\end{equation*}
	where ${\tau}_i^{}=\tfrac{{\tau}}{{M}_i^{}}$ is the learning rate for the $i$-th coordinate. The least squares and logistic regression losses and $\ell_1^{}$  and $\ell_2^{}$ regularizations are considered in this article.
	
	Algorithm \ref{alg:dpcd} summarizes the steps involved in DP-CD. We perform $L$ batches of coordinate descents. In order to guarantee DP, the update to each coordinate is perturbed with noise. The available privacy resource $\eta$ is split equally between $L$ batches as in \cite{mangold2022differentially}. For improved DP-CD,  this $\frac{1}{L}$-th fraction of the resource for each batch gets unequally divided between the coordinate updates (in lines \ref{alg:steps:inner_loop_start}-\ref{alg:steps:inner_loop_end}) to harness the disparity (see  Section \ref{sec:resource_allocation}). The gradients have to be bounded to calibrate the noise; the $i$-th coordinate gradients corresponding to each user are clipped to have a maximum absolute value of $C_i^{}$ and averaged. Hence, the sensitivity of the $i$-th coordinate update is ${\lambda}_i^{}=2\psp\frac{{\tau}_i^{}\psp C_i^{}}{N}$. The clipping constants are adaptively chosen as $C_i^{}=C\sqrt{\frac{{M}_i^{}}{\sum_{j=1}^{K} {M}_j^{}}}$. The hyperparameters $L, {\tau}$ and ${C}$ are tuned as described in \cite{mangold2022differentially}. Also, we use the prescribed values for all other hyperparameters from \cite{mangold2022differentially}.
	\begin{algorithm}[h]
		\caption{Improved Differentially Private Coordinate Descent (Improved DP-CD).}
		\label{alg:dpcd}
		\textbf{Input}: Dataset $\mathcal{D}$, privacy parameters $\epsilon\in\mathbb{R}_{++}^{}$ and $\delta\in(0,1)$, iteration budget $L\in\mathbb{N}$, initial point $\mbf{{\theta}}^{(0)}_{}\in \mathbb{R}^K_{}$, Clipping constants $\{{C}_i^{}\}_{i=1}^K$, and step sizes $\{{\tau}_i^{}\}_{i=1}^K\pspp$.
		\begin{algorithmic}[1]
			\State Determine Gaussian noise scales $\{\sigma_i^{}\}_{i=1}^K$  from the privacy parameters.
			\For{$l\in \mathbb{N}_{L}^{}$}
			\State $\mbf{{\theta}}^{(l)}_{} \gets \mbf{{\theta}}^{(l-1)}_{}\pspp$. 
			\For{$i\in \mathbb{N}_{K}^{}$}
			\State \label{alg:steps:inner_loop_start}
			Sample $t_i^{(l)}\sim \mathcal{N}(0,\sigma_j^2)\pspp$.
			\State ${\vartheta}^{(l)}_i=\frac{1}{n}\nspp\sum_{n=1}^N \operatorname{clip}_{}^{}\left(\nabla_i^{}\pspp J(\mbf{{\theta}}_{}^{(l)}; \mathcal{D}_n^{});\pspp 
			C_i^{}\right)\pspp$.
			\State \label{alg:steps:inner_loop_end}
			${\theta}^{(l)}_i \gets \operatorname{prox}_{{\tau}_i^{}{\psi}_i^{}}^{} \!\nspp \left( {\theta}^{(l)}_i- {\tau}_i^{} {\vartheta}^{(l)}_i +t^{(l)}_i \right)\pspp$.
			\EndFor
			\EndFor
		\end{algorithmic}
		\textbf{Output}: $\widehat{\mbf{{\theta}}} = \mbf{{\theta}}^{(L)}_{}$.
	\end{algorithm}

	\subsection*{Private estimation of smoothness constants}
	
	For DP-CD, the sensitivity profile is determined by the coordinate-wise smoothness constants of the objective function, $\{{M}_i^{}\}_{i=1}^{K}$ \cite{mangold2022differentially}. As described above, the coordinate-wise learning rates are set as ${\tau}_i^{}=\tfrac{{\tau}}{{M}_i^{}}\pspp\nspp$, and the clipping constants are taken as 
	\begin{equation*}
		C_i^{}=C\sqrt{\frac{{M}_i^{}}{\sum_{j=1}^{K} {M}_j^{}}}, 
		\,\ i\in \mathbb{N}_{K}^{}
		\pspp\nspp,  
	\end{equation*}
	where $\tau$ and $C$ are tunable hyperparameters. Thus, the sensitivity profile is given by 
	\begin{equation*}
		{\lambda}_i^{}=2\psp\frac{{\tau}_i^{}\psp C_i^{}}{N}=2\psp\frac{{\tau}\psp C}{N}\bigg( M_i^{} \summ_{j=1}^{K} {M}_j^{}\bigg)^{\nsp-\frac{1}{2}}_{}
		\nsp, 
		\,\ i\in \mathbb{N}_{K}^{}
		\pspp\nspp.
	\end{equation*}
	
	Smoothness constants essentially depend on data, and hence, whenever they are unavailable, they must be estimated privately using a fraction of the available privacy resources. For generalized linear models and their regularized variants (like LASSO and logistic regression with $\ell_2^{}$-regularization considered in  Section \ref{sec:appn_dpcd}), smoothness constants can be easily estimated from data: For these problems, we have $M_i^{}\propto \frac{1}{N}\sum_{n=1}^{N} x_{n,\pspp i}^{2}\pspp$, where $x_{n,\pspp i}^{}$	is the $i$-th attribute of the $n$-th user, and thus, private estimation of $\{{M}_i^{}\}_{i=1}^{K}$ boils down to the classic private mean estimation problem.
	
	We consider that a portion $\epsilon'_{}$ of the available $\epsilon$ be used to estimate the smoothness constants as explained in \cite[Appx. F]{mangold2022differentially}. As briefed above, for generalized models, $M_i^{}$ is given by the average of $M_i^{(n)}=	x_{n,\pspp i}^{2}$ over $n\in\mathbb{N}_N^{}$. With the knowledge of approximate upper bound $b_i^{}$ on the magnitude of the $i$-th feature, we can obtain the private smoothness constant ${M}^{}_{i,\pspp\texttt{priv}}$ by perturbing the average of clipped $M_i^{(n)}$'s with Laplace noise; with the budget of $\frac{\epsilon'}{K}$ to estimate each smoothness constant, we have
	\begin{equation*}
		{M}^{}_{i,\pspp\texttt{priv}}=\frac{1}{N}\sum_{i=1}^N\operatorname{clip}\big(M_i^{(n)};\pspp b_i^{2}\big)
		+
		\frac{b_i^2\pspp K}{N\pspp \epsilon'}\pspp T_i^{}
		\pspp,
		\quad
		i\in \mathbb{N}_{K}^{}
		\pspp,
	\end{equation*}
	where $T_i^{}$ are independent standard Laplace variables, $T_i^{}\sim \mathcal{L}(0,1)\pspp$. These privately estimated constants can then be used directly in DP-CD, using the remaining budget of $\epsilon-\epsilon'$ for the optimization procedure.
	
	For the experiments considered in Section \ref{sec:appn_dpcd}, we now present the results when a part of the available $\epsilon$ is used for estimating the sensitivity profile as outlined above. Table \ref{tab:dp_cd} provides the relative error to the (non-private) optimal objective value under DP-CD and improved DP-CD for different values of $\frac{\epsilon'}{\epsilon}$. It can be observed that when the fraction of budget used to estimate the smoothness constants is around $10-15$\%, the performance of both DP-CD and improved DP-CD does not deteriorate much compared to the setting where $\{{M}_i^{}\}_{i=1}^{K}$ are given \textit{a priori}. This observation concurs with those in \cite{mangold2022differentially}. For instance, DP-CD and improved DP-CD, respectively, results in an excess relative error of only $0.018$\% and $0.03$\% for the LASSO problem on the California dataset, and for the logistic regression with $\ell_2^{}$-regularization on the Electricity dataset, the respective deteriorations are $0.033$\% and $0.016$\%.	As with our previous results, we observe that the improved DP-CD adopting the proposed i.n.i.d. mechanism performs better than DP-CD even in this setting. 
	\begin{table}[h!]
		\begin{center}
			\caption{Relative error (in \%) to the optimal objective value in DP-CD with private estimation of smoothness constants.}
			\label{tab:dp_cd}
			\setlength{\tabcolsep}{4pt}
			\begin{tabular}{? C{1.5em} | C{4em} ? C{3.5em}? *{4}{C{2.7em}|} C{2.7em}?}
				\clineB{3-8}{2}
				\multicolumn{1}{c}{} &\multicolumn{1}{c?}{} & 
				\multirow{2}{*}{
					\begin{tabular}{c}
						\\[-2em]
						\hspace{-1.05em}\textbf{$\{{M}_i^{}\}_{i=1}^{K}$}
						\\[-0.5em]
						\hspace{-1.5em} \textbf{available}
					\end{tabular}				
				} 	
				& \multicolumn{5}{c?}{\textbf{Fraction of budget to estimate}  $\{{M}_i^{}\}_{i=1}^{K}$}  	\\[0.35em] 
				\cline{4-8}
				\multicolumn{1}{c}{} &\multicolumn{1}{c?}{} & 	& \textbf{5\%} 	& \textbf{10\%} 	& \textbf{15\%} 	& \textbf{20\%} 	& \textbf{25\%} 	\\[0.35em] 
				\clineB{3-8}{2}
				\specialrule{1pt}{2pt}{0pt}
				
				\multirow{2}{*}{\spheading{\centering\begin{tabular}{c}
							\\[-3.05em]
							\hspace{-0.45em}\textbf{California}
							\\[-1em]
							\hspace{-0.9em} \textbf{dataset}
						\end{tabular}$\quad\;\ \ $}}
				& \textbf{DP-CD} & 1.8836 	& 2.9061 	& \textbf{1.9013} 	& 2.4087 	& 2.7861 	& 3.7214 	\\[0.75em]  
				\cline{2-8}
				
				& \begin{tabular}{c}
					\\[-2.5em]
					\hspace{-0.575em}\textbf{Improved}
					\\[-0.75em]
					\hspace{-1em} \textbf{DP-CD}
				\end{tabular} & 0.4469 	& 1.2029 	& 0.5250 	& \textbf{0.4768} 	& 1.1441 	& 1.2747 	\\[0.3em]  
				\clineB{1-8}{2}
				\specialrule{1pt}{2pt}{0pt}	
				
				\multirow{2}{*}{\spheading{\centering\begin{tabular}{c}
							\\[-3.05em]
							\hspace{-0.5em}\textbf{Electricity}
							\\[-1em]
							\hspace{-0.95em} \textbf{dataset}
						\end{tabular}$\quad\;\ \ $}}
				& \textbf{DP-CD} & 0.2157 	& 0.2934 	& 0.2528 	& \textbf{0.2491} 	& 0.2614 	& 0.2838 	\\[0.75em]  
				\cline{2-8}
				
				& \begin{tabular}{c}
					\\[-2.5em]
					\hspace{-0.575em}\textbf{Improved}
					\\[-0.75em]
					\hspace{-1em} \textbf{DP-CD}
				\end{tabular} & 0.1263 	& 0.1443 	& \textbf{0.1429} 	& 0.1651 	& 0.1979 	& 0.2186 	\\[0.3em] 
				\clineB{1-8}{2}
				\specialrule{1pt}{0pt}{1pt}
			\end{tabular}
		\end{center}
	\end{table}
	
	 \section*{Acknowledgment}
	 \addcontentsline{toc}{section}{Acknowledgment}
	 We thank the anonymous reviewers and the associate editor for their constructive feedback and suggestions, which helped to improve the article. Special thanks to Janani Suresh, MS scholar at IIT Madras, for her help with the simulations.

	{
		\bibliographystyle{IEEEtran}
		\bibliography{ref_inid}

\begin{thebibliography}{10}
\providecommand{\url}[1]{#1}
\csname url@samestyle\endcsname
\providecommand{\newblock}{\relax}
\providecommand{\bibinfo}[2]{#2}
\providecommand{\BIBentrySTDinterwordspacing}{\spaceskip=0pt\relax}
\providecommand{\BIBentryALTinterwordstretchfactor}{4}
\providecommand{\BIBentryALTinterwordspacing}{\spaceskip=\fontdimen2\font plus
\BIBentryALTinterwordstretchfactor\fontdimen3\font minus
  \fontdimen4\font\relax}
\providecommand{\BIBforeignlanguage}[2]{{%
\expandafter\ifx\csname l@#1\endcsname\relax
\typeout{** WARNING: IEEEtran.bst: No hyphenation pattern has been}%
\typeout{** loaded for the language `#1'. Using the pattern for}%
\typeout{** the default language instead.}%
\else
\language=\csname l@#1\endcsname
\fi
#2}}
\providecommand{\BIBdecl}{\relax}
\BIBdecl

\bibitem{dwork2014algorithmic}
C.~Dwork and A.~Roth, ``The algorithmic foundations of differential privacy.''
  \emph{Found. Trends Theor. Comput. Sci.}, vol.~9, no. 3-4, pp. 211--407,
  2014.

\bibitem{wang2018revisiting}
Y.-X. Wang, ``Revisiting differentially private linear regression: optimal and
  adaptive prediction \& estimation in unbounded domain,'' in \emph{Uncertainty
  in Artif. Intell.}, 2018.

\bibitem{chaudhuri2011differentially}
K.~Chaudhuri, C.~Monteleoni, and A.~D. Sarwate, ``Differentially private
  empirical risk minimization.'' \emph{J. Mach. Learn. Res.}, vol.~12, no.~3,
  2011.

\bibitem{dwork2014analyze}
C.~Dwork, K.~Talwar, A.~Thakurta, and L.~Zhang, ``Analyze {G}auss: optimal
  bounds for privacy-preserving principal component analysis,'' in \emph{Proc.
  Annu. ACM Symp. Theory of Comput.}, 2014, pp. 11--20.

\bibitem{shechner2020private}
M.~Shechner, O.~Sheffet, and U.~Stemmer, ``Private $k$-means clustering with
  stability assumptions,'' in \emph{Proc. Int. Conf. Artif. Intell. and
  Statist.}\hskip 1em plus 0.5em minus 0.4em\relax PMLR, 2020, pp. 2518--2528.

\bibitem{us2021disclosure}
{US Census Bureau}, ``Disclosure avoidance for the 2020 census: an
  introduction,'' 2021.

\bibitem{dwork2006our}
C.~Dwork, K.~Kenthapadi, F.~McSherry, I.~Mironov, and M.~Naor, ``Our data,
  ourselves: Privacy via distributed noise generation,'' in \emph{Proc. Annu.
  Int. Conf. Theory Appl. Cryptograph. Techn.}\hskip 1em plus 0.5em minus
  0.4em\relax Springer, 2006, pp. 486--503.

\bibitem{le2013differentially}
J.~Le~Ny and G.~J. Pappas, ``Differentially private filtering,'' \emph{IEEE
  Trans. Automatic Control}, vol.~59, no.~2, pp. 341--354, 2013.

\bibitem{balle2018improving}
B.~Balle and Y.-X. Wang, ``Improving the {G}aussian mechanism for differential
  privacy: Analytical calibration and optimal denoising,'' in \emph{Proc. Int.
  Conf. Mach. Learn.}\hskip 1em plus 0.5em minus 0.4em\relax PMLR, 2018, pp.
  394--403.

\bibitem{dong2022gaussian}
J.~Dong, A.~Roth, and W.~J. Su, ``Gaussian differential privacy,'' \emph{J.
  Roy. Statistical Soc.: Series B}, vol.~84, no.~1, pp. 3--37, 2022.

\bibitem{steinke2022composition}
T.~Steinke, ``Composition of differential privacy \& privacy amplification by
  subsampling,'' \emph{arXiv:2210.00597}, 2022.

\bibitem{liu2018generalized}
F.~Liu, ``Generalized {G}aussian mechanism for differential privacy,''
  \emph{IEEE Trans. Knowledge and Data Engg.}, vol.~31, no.~4, pp. 747--756,
  2018.

\bibitem{sadeghi2022offset}
P.~Sadeghi and M.~Korki, ``Offset-symmetric {G}aussians for differential
  privacy,'' \emph{IEEE Trans. Inf. Forensics Security}, vol.~17, pp.
  2394--2409, 2022.

\bibitem{muthukrishnan2023grafting}
G.~Muthukrishnan and S.~Kalyani, ``Grafting {L}aplace and {G}aussian
  {D}istributions: {A} {N}ew {N}oise {M}echanism for {D}ifferential
  {P}rivacy,'' \emph{IEEE Trans. Inf. Forensics Security}, vol.~18, pp.
  5359--5374, 2023.

\bibitem{geng2016optimalstaircase}
Q.~Geng and P.~Viswanath, ``The optimal noise-adding mechanism in differential
  privacy,'' \emph{IEEE Trans. Inf. Theory}, vol.~62, no.~2, pp. 925--951,
  2016.

\bibitem{geng2020tighttrunclapl}
Q.~\hspace{0em}Geng, W.~Ding, R.~Guo, and S.~Kumar, ``Tight analysis of privacy
  and utility tradeoff in approximate differential privacy,'' in \emph{Proc.
  Int. Conf. Artif. Intell. Statist.}\hskip 1em plus 0.5em minus 0.4em\relax
  PMLR, 2020, pp. 89--99.

\bibitem{geng2016optimaluniflap}
Q.~Geng and P.~\hspace{0em}Viswanath, ``Optimal noise adding mechanisms for
  approximate differential privacy,'' \emph{IEEE Trans. Inf. Theory}, vol.~62,
  no.~2, pp. 952--969, 2016.

\bibitem{geng2015staircase}
Q.~Geng, P.~Kairouz, S.~Oh, and P.~Viswanath, ``The staircase mechanism in
  differential privacy,'' \emph{IEEE J. Sel. Topics Signal Process.}, vol.~9,
  no.~7, pp. 1176--1184, 2015.

\bibitem{vinterbo2022differential}
S.~A. Vinterbo, ``Differential privacy for symmetric log-concave mechanisms,''
  in \emph{Proc. Int. Conf. Artif. Intell. and Statist.}\hskip 1em plus 0.5em
  minus 0.4em\relax PMLR, 2022, pp. 6270--6291.

\bibitem{dong2021central}
J.~Dong, W.~Su, and L.~Zhang, ``A central limit theorem for differentially
  private query answering,'' in \emph{Proc. Adv. Neural Inf. Process. Syst.},
  vol.~34.\hskip 1em plus 0.5em minus 0.4em\relax PMLR, 2021, pp.
  14\,759--14\,770.

\bibitem{sommer2019privacy}
D.~M. Sommer, S.~Meiser, and E.~Mohammadi, ``Privacy loss classes: The central
  limit theorem in differential privacy,'' \emph{Proc. Privacy Enhancing
  Technologies}, vol.~2, pp. 245--269, 2019.

\bibitem{mangold2022differentially}
P.~Mangold, A.~Bellet, J.~Salmon, and M.~Tommasi, ``Differentially private
  coordinate descent for composite empirical risk minimization,'' in
  \emph{Proc. Int. Conf. Mach. Learn.}\hskip 1em plus 0.5em minus 0.4em\relax
  PMLR, 2022, pp. 14\,948--14\,978.

\bibitem{sander2023tan}
T.~Sander, P.~Stock, and A.~Sablayrolles, ``{TAN} without a burn: Scaling laws
  of {DP-SGD},'' in \emph{Proc. Int. Conf. Mach. Learn.}\hskip 1em plus 0.5em
  minus 0.4em\relax PMLR, 2023, pp. 29\,937--29\,949.

\bibitem{li2010optimizing}
C.~Li, M.~Hay, V.~Rastogi, G.~Miklau, and A.~McGregor, ``Optimizing linear
  counting queries under differential privacy,'' in \emph{Proc. ACM
  SIGMOD-SIGACT-SIGART Sym. Principles of Database Syst.}, 2010, pp. 123--134.

\bibitem{nikolov2013geometry}
A.~Nikolov, K.~Talwar, and L.~Zhang, ``The geometry of differential privacy:
  the sparse and approximate cases,'' in \emph{Proc. Annual ACM Symp. Theory of
  Computing}, 2013, pp. 351--360.

\bibitem{edmonds2020power}
A.~Edmonds, A.~Nikolov, and J.~Ullman, ``The power of factorization mechanisms
  in local and central differential privacy,'' in \emph{Proc. Annual ACM Symp.
  Theory of Computing}, 2020, pp. 425--438.

\bibitem{hardt2010geometry}
M.~Hardt and K.~Talwar, ``On the geometry of differential privacy,'' in
  \emph{Proceedings of the forty-second ACM symposium on Theory of computing},
  2010, pp. 705--714.

\bibitem{awan2021structure}
J.~Awan and A.~Slavkovi{\'c}, ``Structure and sensitivity in differential
  privacy: Comparing {$K$}-norm mechanisms,'' \emph{J. Amer. Stat. Assoc.},
  vol. 116, no. 534, pp. 935--954, 2021.

\bibitem{gaboardi2016psi}
M.~Gaboardi, J.~Honaker, G.~King, J.~Murtagh, K.~Nissim, J.~Ullman, and
  S.~Vadhan, ``Psi ($\psi$): a private data sharing interface,'' \emph{arXiv
  preprint arXiv:1609.04340}, 2016.

\bibitem{ryu2024noise}
S.~Ryu, J.~Jang, and H.~J. Yang, ``Noise variance optimization in differential
  privacy: A game-theoretic approach through per-instance differential
  privacy,'' \emph{IEEE Access}, vol.~12, pp. 103\,104--103\,118, 2024.

\bibitem{bun2016concentrated}
M.~Bun and T.~Steinke, ``Concentrated differential privacy: Simplifications,
  extensions, and lower bounds,'' in \emph{Proc. Int. Conf. Theory of Cryptogr.
  Part I}.\hskip 1em plus 0.5em minus 0.4em\relax Springer, 2016, pp. 635--658.

\bibitem{hogg2019introduction}
R.~V. Hogg and A.~T. Craig, \emph{Introduction to Mathematical Statistics},
  8th~ed.\hskip 1em plus 0.5em minus 0.4em\relax Pearson Education, Inc., 2019.

\bibitem{mironov2017renyi}
I.~Mironov, ``R{\'e}nyi differential privacy,'' in \emph{Proc. IEEE Comput.
  Secur. Found. Symp.}\hskip 1em plus 0.5em minus 0.4em\relax IEEE, 2017, pp.
  263--275.

\bibitem{abadi2016deep}
M.~Abadi, A.~Chu, I.~Goodfellow, H.~B. McMahan, I.~Mironov, K.~Talwar, and
  L.~Zhang, ``Deep learning with differential privacy,'' in \emph{Proc. ACM
  SIGSAC Conf. Computer and Communications security}, 2016, pp. 308--318.

\bibitem{wang2019subsampled}
Y.-X. Wang, B.~Balle, and S.~P. Kasiviswanathan, ``Subsampled {R{\'e}nyi}
  differential privacy and analytical moments accountant,'' in \emph{Proc. Int.
  Conf. Artif. Intell. and Statist.}\hskip 1em plus 0.5em minus 0.4em\relax
  PMLR, 2019, pp. 1226--1235.

\bibitem{zhu2022optimal}
Y.~Zhu, J.~Dong, and Y.-X. Wang, ``Optimal accounting of differential privacy
  via characteristic function,'' in \emph{Proc. Int. Conf. Artif. Intell. and
  Statist.}\hskip 1em plus 0.5em minus 0.4em\relax PMLR, 2022, pp. 4782--4817.

\bibitem{asoodeh2020better}
S.~Asoodeh, J.~Liao, F.~P. Calmon, O.~Kosut, and L.~Sankar, ``A better bound
  gives a hundred rounds: Enhanced privacy guarantees via f-divergences,'' in
  \emph{IEEE Int. Symp. Inf. Theory}.\hskip 1em plus 0.5em minus 0.4em\relax
  IEEE, 2020, pp. 920--925.

\bibitem{balle2020privacy}
B.~Balle, G.~Barthe, and M.~Gaboardi, ``Privacy profiles and amplification by
  subsampling,'' \emph{J. Privacy and Confidentiality}, vol.~10, no.~1, 2020.

\bibitem{goldberg1987equivalence}
M.~Goldberg, ``Equivalence constants for $l_p^{}$ norms of matrices,''
  \emph{Linear and Multilinear Algebra}, vol.~21, no.~2, pp. 173--179, 1987.

\bibitem{zhang2012functional}
J.~Zhang, Z.~Zhang, X.~Xiao, Y.~Yang, and M.~Winslett, ``Functional mechanism:
  Regression analysis under differential privacy,'' \emph{Proc. VLDB
  Endowment}, vol.~5, no.~11, 2012.

\bibitem{yu2014differentially}
F.~Yu, M.~Rybar, C.~Uhler, and S.~E. Fienberg, ``Differentially-private
  logistic regression for detecting multiple-snp association in gwas
  databases,'' in \emph{Proc. Int. Conf. Privacy in Statistical Databases:
  UNESCO Chair in Data Privacy}.\hskip 1em plus 0.5em minus 0.4em\relax
  Springer, 2014, pp. 170--184.

\bibitem{tian2018selective}
X.~Tian and J.~Taylor, ``Selective inference with a randomized response,''
  \emph{Ann. Statist.}, vol.~46, no.~2, pp. 679--710, 2018.

\bibitem{dwork2006calibrating}
C.~Dwork, F.~McSherry, K.~Nissim, and A.~Smith, ``Calibrating noise to
  sensitivity in private data analysis,'' in \emph{Proc. Theory Cryptogr.
  Conf.}\hskip 1em plus 0.5em minus 0.4em\relax Springer, 2006, pp. 265--284.

\bibitem{burden2011numerical}
R.~L. Burden and J.~D. Faires, \emph{Numerical Analysis}.\hskip 1em plus 0.5em
  minus 0.4em\relax Brooks/Cole, Cengage Learning, 2011.

\bibitem{steele2004cauchy}
J.~M. Steele, \emph{The Cauchy-Schwarz master class: an introduction to the art
  of mathematical inequalities}.\hskip 1em plus 0.5em minus 0.4em\relax
  Cambridge Univ. Press, 2004.

\bibitem{he2022exploring}
J.~He, X.~Li, D.~Yu, H.~Zhang, J.~Kulkarni, Y.~T. Lee, A.~Backurs, N.~Yu, and
  J.~Bian, ``Exploring the limits of differentially private deep learning with
  group-wise clipping,'' in \emph{Proc. Int. Conf. Learn. Representations},
  2022.

\bibitem{marshall2011inequalities}
A.~W. Marshall, I.~Olkin, and B.~C. Arnold, \emph{Inequalities: theory of
  majorization and its applications}.\hskip 1em plus 0.5em minus 0.4em\relax
  Springer, 2011.

\bibitem{hardy1929some}
G.~H. Hardy, J.~E. Littlewood, and G.~P\'{o}lya, ``Some simple inequalities
  satisfied by convex functions,'' \emph{Messenger Math.}, vol.~58, pp.
  145--152, 1929.

\bibitem{geng2015staircasearxiv}
Q.~\hspace{0em}Geng and P.~Viswanath, ``The optimal mechanism in differential
  privacy: Multidimensional setting,'' \emph{arXiv:1312.0655}, 2013.

\bibitem{californiadataset}
R.~K. Pace and R.~Barry, ``Sparse spatial autoregressions,'' \emph{Statist.
  Probability Lett.}, vol.~33, no.~3, pp. 291--297, 1997.

\bibitem{electricitydataset}
\BIBentryALTinterwordspacing
Electricity dataset. [Online]. Available: \url{https://www.openml.org/d/151}
\BIBentrySTDinterwordspacing

\bibitem{mangoubi2022re}
O.~Mangoubi and N.~Vishnoi, ``Re-analyze gauss: Bounds for private matrix
  approximation via dyson brownian motion,'' vol.~35, pp. 38\,585--38\,599,
  2022.

\bibitem{rahmani2017coherence}
M.~Rahmani and G.~K. Atia, ``Coherence pursuit: Fast, simple, and robust
  principal component analysis,'' \emph{IEEE Trans. Signal Process.}, vol.~65,
  no.~23, pp. 6260--6275, 2017.

\bibitem{soltanolkotabi2012geometric}
M.~Soltanolkotabi and E.~J. Cand{\'e}s, ``A geometric analysis of subspace
  clustering with outliers,'' \emph{Ann. Statist.}, vol.~40, no.~4, pp.
  2195--2238, 2012.

\bibitem{bassily2014private}
R.~Bassily, A.~Smith, and A.~Thakurta, ``Private empirical risk minimization:
  Efficient algorithms and tight error bounds,'' in \emph{Proc. IEEE Annu.
  Symp. Found. Comput. Sci.}\hskip 1em plus 0.5em minus 0.4em\relax IEEE, 2014,
  pp. 464--473.

\bibitem{zhu2019poission}
Y.~Zhu and Y.-X. Wang, ``{Poission subsampled R{\'e}nyi differential
  privacy},'' in \emph{Proc. Int. Conf. Mach. Learn.}\hskip 1em plus 0.5em
  minus 0.4em\relax PMLR, 2019, pp. 7634--7642.

\bibitem{mironov2019renyi}
I.~Mironov, K.~Talwar, and L.~Zhang, ``R\'{e}nyi differential privacy of the
  sampled {Gaussian} mechanism,'' \emph{arXiv:1908.10530}, 2019.

\bibitem{mcmahan2018learning}
H.~B. McMahan, D.~Ramage, K.~Talwar, and L.~Zhang, ``Learning differentially
  private recurrent language models,'' in \emph{Proc. Int. Conf. Learn.
  Representations}, 2018.

\bibitem{mcmahan2018general}
H.~B. McMahan, G.~Andrew, U.~Erlingsson, S.~Chien, I.~Mironov, N.~Papernot, and
  P.~Kairouz, ``A general approach to adding differential privacy to iterative
  training procedures,'' \emph{arXiv:1812.06210}, 2018.

\bibitem{xiao2023geometry}
H.~Xiao, J.~Wan, and S.~Devadas, ``Geometry of sensitivity: Twice sampling and
  hybrid clipping in differential privacy with optimal gaussian noise and
  application to deep learning,'' in \emph{Proc. ACM SIGSAC Conf. Computer and
  Communications security}, 2023, pp. 2636--2650.

\bibitem{koskela2020computing}
A.~Koskela, J.~J{\"a}lk{\"o}, and A.~Honkela, ``Computing tight differential
  privacy guarantees using {FFT},'' in \emph{Proc. Int. Conf. Artif. Intell.
  and Statist.}\hskip 1em plus 0.5em minus 0.4em\relax PMLR, 2020, pp.
  2560--2569.

\bibitem{boyd2004convex}
S.~P. Boyd and L.~Vandenberghe, \emph{Convex optimization}.\hskip 1em plus
  0.5em minus 0.4em\relax Cambridge Univ. Press, 2004.

\end{thebibliography}
	}
	
\end{document}